\RequirePackage{fix-cm} 
\documentclass[a4paper, twoside, reqno, dvips, 12pt]{amsart}
\usepackage{fixltx2e}   

\usepackage{etex}

\usepackage[latin1]{inputenc}
\usepackage[T1]{fontenc}

\usepackage{esint}
\usepackage{dsfont}
\usepackage{xspace}
\usepackage{amsgen}
\usepackage{amsthm}
\usepackage{amssymb}
\usepackage{amsmath}
\usepackage{wasysym}
\usepackage{upgreek}
\usepackage{amsfonts}
\usepackage{stmaryrd}
\usepackage{scalerel}
\usepackage{mathtools}
\usepackage{stackengine}

\usepackage{relsize}
\usepackage[mathcal, mathscr]{euscript}

\usepackage{mathrsfs}
\DeclareMathAlphabet{\mathscrbf}{OMS}{mdugm}{b}{n}

\usepackage{a4wide}

\usepackage[dvipsnames, table]{xcolor}
\definecolor{bckg}{RGB}{20.8, 20.8, 20.8}
\definecolor{oneblue}{rgb}{0.0, 0.0, 0.85}
\definecolor{Lightblue}{RGB}{214, 214, 214}
\definecolor{bluepigment}{rgb}{0.2, 0.2, 0.6}
\definecolor{charcoal}{rgb}{0.21, 0.27, 0.31}
\definecolor{denimblue}{rgb}{0.08, 0.38, 0.74}
\definecolor{Lightgray}{rgb}{0.89, 0.89, 0.89}
\definecolor{darkgrey}{rgb}{0.273, 0.281, 0.30}
\definecolor{darkelectricblue}{rgb}{0.33, 0.41, 0.47}

\usepackage[sort&compress, comma, square, numbers]{natbib}

\usepackage{psfrag}
\usepackage{graphicx}
\usepackage{adjustbox}
\usepackage[tight]{subfigure}
\usepackage{morefloats}
\usepackage{indentfirst}

\usepackage[usenames, dvipsnames, pdf]{pstricks}
\usepackage{epsfig}
\usepackage{pst-grad} 
\usepackage{pst-plot} 

\usepackage{acronym}
\usepackage{microtype}
\usepackage[labelsep=period,%
            labelfont={bf,sf,color=bluepigment},%
            justification=raggedright]{caption}

\usepackage[perpage, symbol]{footmisc}

\usepackage[colorlinks,
           urlcolor=oneblue,
           linkcolor=denimblue,
           citecolor=NavyBlue,
           bookmarksopen=false,
           pdfpagemode=UseNone,
           pagebackref]{hyperref}

\usepackage[explicit]{titlesec}

\titleformat{\section}[block]
  {\color{NavyBlue}\Large\sffamily\bfseries}
  {}
  {0.0em}
  {\colorbox{bckg!5}{\strut\parbox{\dimexpr\linewidth-2\fboxsep\relax}{\thesection. #1}}}
  [\vspace*{0.33em}]

\titleformat{name=\section,numberless}[block]
  {\color{NavyBlue}\Large\sffamily\bfseries}
  {}
  {0.0em}
  {\colorbox{bckg!5}{\strut\parbox{\dimexpr\linewidth-2\fboxsep\relax}{#1}}}
  [\vspace*{0.33em}]

\titleformat{\subsection}
  {\color{NavyBlue}\large\sffamily\bfseries}
  {}
  {0.0em}
  {\colorbox{bckg!5}{\parbox{\dimexpr\linewidth-2\fboxsep\relax}{\thesubsection. #1}}}
  [\vspace*{0.33em}]

\titleformat{name=\subsection,numberless}
  {\color{NavyBlue}\Large\sffamily\bfseries}
  {}
  {0em}
  {\colorbox{bckg!5}{\parbox{\dimexpr\linewidth-2\fboxsep\relax}{#1}}}
  [\vspace*{0.33em}]

\titleformat{\subsubsection}
  {\color{bluepigment}\sffamily\normalsize\bfseries}
  {\thesubsubsection}
  {0.5em}
  {#1}
  [\vspace*{0.33em}]

\titleformat{\paragraph}[runin]
  {\color{bluepigment}\sffamily\small\bfseries}
  {}
  {0em}
  {#1}

\titlespacing{\section}{0.0em}{1.5em plus 2pt minus 2pt}%
{1.0em plus 2pt minus 2pt}[0em]
\titlespacing{\subsection}{0.5em}{1.5em plus 2pt minus 2pt}%
{1.0em}[0em]
\titlespacing{\subsubsection}{0.5em}{1.5em plus 2pt minus 2pt}%
{1.0em plus 2pt minus 2pt}[0em]

\usepackage{titletoc}

\contentsmargin{0.5em}
\newlength{\tocsep} 
\titlecontents{section}[\tocsep]
  {\addvspace{10pt}\bfseries\sffamily}
  {\contentslabel[\thecontentslabel]{\tocsep}}
  {}
  {\ \titlerule*[0.75pc]{.}\ \thecontentspage}
  []
\titlecontents{subsection}[\tocsep]
  {\addvspace{8pt}\sffamily}
  {\contentslabel[\thecontentslabel]{\tocsep}}
  {}
  {\ \titlerule*[0.5pc]{.}\ \thecontentspage}
  []
\titlecontents*{subsubsection}[\tocsep]
  {\addvspace{2pt}\footnotesize\sffamily}
  {}
  {}
  {\ \titlerule*[0.35pc]{.}\ \thecontentspage}
  [\\*]

\makeatletter
\def\@setauthors{%
  \begingroup
  \def\thanks{\protect\thanks@warning}%
  \trivlist
  \centering\footnotesize \@topsep30\p@\relax
  \advance\@topsep by -\baselineskip
  \item\relax
  \author@andify\authors
  \def\\{\protect\linebreak}%
  \textsc{\normalsize\textcolor{darkelectricblue}{\authors}}%
  \ifx\@empty\contribs
  \else
    ,\penalty-3 \space \@setcontribs
    \@closetoccontribs
  \fi
  \endtrivlist
  \endgroup
}
\def\@settitle{\begin{center}%
  \baselineskip14\p@\relax
    \bfseries
    \textsc{\Large\textcolor{charcoal}{\@title}}
  \end{center}%
}
\makeatother

\usepackage{enumitem}
\setlist[description]{%
  topsep=30pt,               
  itemsep=5pt,               
  font={\bfseries\sffamily\color{NavyBlue}}, 
}

\usepackage{fancyhdr}
\usepackage{lastpage}

\newcommand*\Title{\textcolor{bluepigment}{Singular solitary waves}}
\newcommand*\Authors{\textcolor{bluepigment}{D.~Clamond, D.~Dutykh \& A.~Galligo}}
\newcommand*{\plogo}{\textcolor{gray}{{\texttt{arXiv.org} / \textsc{hal}}}} 

\numberwithin{equation}{section}

\newcommand{\R}{\mathds{R}}

\newcommand{\ud}{\mathrm{d}}

\renewcommand{\beta}{\upbeta}

\newcommand{\D}{\mathscrbf{D}}

\renewcommand{\alpha}{\upalpha}

\newcommand{\uD}{\mathrm{D}}

\newcommand{\Bo}{\mathsf{Bo}}
\newcommand{\Fr}{\mathsf{Fr}}

\newcommand{\We}{\mathsf{We}}

\renewcommand{\Re}{\operatorname{Re}}

\newcommand{\ie}{\emph{i.e.}\/ }
\newcommand{\eg}{\emph{e.g.}\/ }
\newcommand{\etc}{\emph{etc.}\/}

\renewcommand{\sim}{\thicksim}

\newcommand{\half}{{\textstyle{1\over2}}}
\newcommand{\third}{{\textstyle{1\over3}}}

\usepackage{acronym}
\acrodef{bvp}[BVP]{Boundary Value Problem}
\acrodef{NSWE}{Nonlinear Shallow Water Equations}

\begin{document}

\title[\Title]{Algebraic method for constructing singular steady solitary waves: A case study}

\author[D.~Clamond]{Didier Clamond$^*$}
\address{Laboratoire J. A. Dieudonn\'e, Universit\'e de Nice -- Sophia Antipolis, Parc Valrose, 06108 Nice cedex 2, France}
\email{diderc@unice.fr}
\urladdr{http://math.unice.fr/~didierc/}
\thanks{$^*$ Corresponding author}

\author[D.~Dutykh]{Denys Dutykh}
\address{LAMA, UMR 5127 CNRS, Universit\'e Savoie Mont Blanc, Campus Scientifique, 
73376 Le Bourget-du-Lac Cedex, France}
\email{Denys.Dutykh@univ-savoie.fr}
\urladdr{http://www.denys-dutykh.com/}

\author[A.~Galligo]{Andr\'e Galligo}
\address{Laboratoire J. A. Dieudonn\'e, Universit\'e de Nice -- Sophia Antipolis, Parc Valrose, 06108 Nice cedex 2, France}
\email{galligo@math.unice.fr}
\urladdr{http://math.unice.fr/~galligo/}



\begin{titlepage}
\thispagestyle{empty} 
\noindent
{\Large Didier \textsc{Clamond}}\\
{\it\textcolor{gray}{Universit\'e de Nice -- Sophia Antipolis, France}}\\[0.02\textheight]
{\Large Denys \textsc{Dutykh}}\\
{\it\textcolor{gray}{CNRS--LAMA, Universit\'e Savoie Mont Blanc, France}}
\\[0.02\textheight]
{\Large Andr\'e \textsc{Galligo}}\\
{\it\textcolor{gray}{Universit\'e de Nice -- Sophia Antipolis, France}}
\\[0.08\textheight]

\vspace*{1.1cm}

\colorbox{Lightblue}{
  \parbox[t]{1.0\textwidth}{
    \centering\huge\sc
    \vspace*{0.7cm}
    
    \textcolor{bluepigment}{Algebraic method for constructing singular steady solitary waves: A case study}
    
    \vspace*{0.7cm}
  }
}

\vfill 

\raggedleft     
{\large \plogo} 
\end{titlepage}


\newpage
\thispagestyle{empty} 
\par\vspace*{\fill}   
\begin{flushright} 
{\textcolor{denimblue}{\textsc{Last modified:}} \today}
\end{flushright}


\newpage
\maketitle
\thispagestyle{empty}


\begin{abstract}

This article describes the use of algebraic methods in a phase plane analysis of ordinary differential equations. The method is illustrated by the study of capillary-gravity steady surface waves propagating in shallow water. We consider the (fully nonlinear, weakly dispersive) Serre--Green--Naghdi equations with surface tension, because it provides a tractable model that, in the same time, is not too simple so the interest of the method can be emphasised. In particular, we analyse a special class of solutions, the solitary waves, which play an important role in many fields of Physics. In capillary-gravity regime, there are two kinds of localised infinitely smooth travelling wave solutions -- solitary waves of elevation and of depression. However, if we allow the solitary waves to have an angular point, the ``zoology'' of solutions becomes much richer and the main goal of this study is to provide a complete classification of such singular localised solutions using the methods of the \emph{effective} Algebraic Geometry.

\bigskip
\noindent \textbf{\keywordsname:} solitary waves, singular solutions, phase plane analysis, algebraic geometry \\


\end{abstract}


\newpage
\tableofcontents
\thispagestyle{empty}


\newpage
\section{Introduction}

Ordinary differential equations (ODE) is a subject of intensive researches in Mathematics and they play a significant role in all fields where Mathematics can be used as a tool to model phenomena (Physics, Chemistry, Biology, Economics, \etc). The solutions of ODEs are of special interest in applications because they allow a qualitative description of the phenomenon modelled by the equation. Unfortunately, analytic solutions can rarely be obtained and, most of the times, only numerical solutions are accessible. However, even numerical solutions can be hard to obtain when the equation is ``\emph{stiff}'' \cite{Hairer1996} or if one is looking for solutions that are not regular. Regular solutions are often the only ones considered in applications, but irregular solutions can be of special interest too. Indeed, such solutions may correspondant to physical phenomena (\eg water waves with angular crest \cite{Stokes1880}) or to relevant approximations of phenomena (\eg shock waves). (Here, we are more specifically interested in nonlinear water waves, but it is not a limitation for the purpose of the present paper.) Irregular solutions have also been proposed as simple analytical models of regular solutions \cite{Dressler1949, Liao2013} and as elementary basis for numerical approximations such as the finite volumes and finite elements methods \cite{LeVeque1992, Mura1992}.

In order to seek for exact solutions or to build approximations of an ODE, a qualitative analysis of the equation is very useful. Such analysis can be performed via the so-called {\em phase plane analysis\/} (PPA) \cite{Jordan2007}. The qualitative study of ODEs is an active research field; the first notions of the theory can be found in \cite{Arnold1996}. The main idea here is to extend the usual PPA, thanks to geometrical tools. Instead of a formal abstract description of the method, we find more enlightening to consider a peculiar example from the physics of nonlinear water waves. Thus, the method is illustrated by the study of capillary-gravity steady surface waves propagating in shallow water. We consider the (fully nonlinear, weakly dispersive) \textsc{Serre--Green--Naghdi} (SGN) equations with surface tension, because it provides a tractable model that, in the same time, is not too simple so the interest of the method can be emphasised.

The general approach to study dynamical systems consists in finding first the equilibria. This part can be done straightforwardly for the SGN system. Our ambition here is to go slightly further and to study the so-called relative equilibria, \ie steady states in a frame of reference moving with the wave. In the water wave community, they are better known as the {\em travelling wave} solutions. They can be divided in two distinct classes: the periodic waves and the solitary waves. It will be shown below (see Section~\ref{secAsymp}) that the SGN model does not admit any generalised solitary waves, which are connected homoclinically to a periodic wave at infinity. In other words, periodic waves do not interact resonantly with solitary waves in this model. Consequently, as the first attempt, in the present paper we focus only on the case of localised solutions, \ie the solitary waves. There exists a huge interest in the scientific community to study this particular class of solutions in shallow water type models, since they might appear as a result of the long time dynamics stemming from a generic initial condition.

For the SGN peculiar application, the method results in studying a family of algebraic implicit curves defined in the phase plane. This family depends on two physical parameters (the Froude $\Fr$ and Bond $\Bo$ numbers) which characterise the relative importance of inertial, gravity and capillary effects. Depending on the values of these parameters one can obtain different flow regimes (\ie sub- or super-critical, for example). For any fixed pair of the parameters $(\Fr,\,\Bo)$, our simple PPA provides qualitative results without constructing the exact (analytic or numerical) solutions to the ODE. In this study, we do not limit ourselves to smooth solutions because the existence of singular solutions to the free surface Euler equations has been known since the celebrated work of G.~G.~\textsc{Stokes} \cite{Stokes1880}, where he proposed an argument towards the existence of the limiting wave. Later peaked-like solutions have been found in numerous approximate models, such as the \textsc{Camassa--Holm} \cite{Camassa1993}, \textsc{Degasperis--Procesi} \cite{Degasperis1999}, just to name a few. Peaked solutions are easily recognisable in the phase plane by symmetric, but discontinuous paths. For the SGN model, we will be able to find all peakon-like solutions with a vanishing second semi-derivative, the latter condition yielding a zero vertical acceleration at the crest as required by the physics.

The continuous dependence on the parameters values applies to singular solutions as well. In other words, a small variation of a generic value of the Froude or Bond numbers (or even both) does not result in any qualitative change of the curve topology in the phase plane as well as in the physical space. The algebraicity of the equations implies that only a finite number of qualitative behaviours (\ie curve topologies) may appear. By applying the appropriate computer algebraic tools (\eg resultants and discriminants), we are able to provide the complete classification and description of possible solution behaviours. It should be noted that in general algebraic polynomial computations are ill-conditioned in the floating point arithmetics. However, the real numbers are present in the model only through the Froude and Bond numbers. By keeping these numbers as symbolic parameters it is possible to perform all the computations in the certified way, which guarantees somehow the qualitative results shown below.

The present manuscript is organised as follows. In Section~\ref{sec:model}, we present the simplified ansatz and explain the derivation of the motion equations. In Section~\ref{sec:steady}, we restrict ourselves to the investigation of steady solitary waves and explain how the problem reduces to the study of a first order nonlinear differential equation. Then, we consider limiting cases. In Section~\ref{secsolexp}, we present our phase plane analysis techniques. The qualitative properties of the solutions and generalised solutions of the studied differential equation are deduced from algebraic-geometric features of a family of plane curves. In Section~\ref{sec:concl}, we summarise our work and propose research directions for future works.


\section{SGN equations with surface tension}
\label{sec:model}

In the present section, we present briefly the derivation of the \textsc{Serre--Green--Naghdi} equations (SGN) in the capillary-gravity regime. For long waves in shallow water, the velocity varies little along the vertical and it is thus mostly horizontal (so-called `columnar flows'). For such flows, a relevant ansatz fulfilling the incompressibility of the fluid and the bottom impermeability is   
\begin{equation}\label{defuvse}
  u(x,y,t)\ \approx\ \bar{u}(x,t), \qquad v(x,y,t)\ \approx\ -\,(y+d)\,\bar{u}_x
\end{equation}
where $d$ is the mean water depth and $\bar{u}$ is the horizontal velocity averaged over the water column --- \ie $\bar{u}\equiv h^{-1}\int_{-d}^\eta u\,\ud y$, $h \equiv \eta + d$ being the total water depth --- $y = \eta$ and $y = 0$ being the equations of the free surface and of the still water level, respectively.

With the ansatz \eqref{defuvse}, the vertical acceleration is
\begin{align}
  \frac{\uD\,v}{\uD\/t}\ &\equiv\ \frac{\partial\,v}{\partial\/t}\ +\ u\,\frac{\partial\,v}{\partial\/x}\ +\ 
  v\,\frac{\partial\,v}{\partial\/y}\ \approx\  -\,v\,\bar{u}_x\ -\ (y+d)\,\frac{\uD\,\bar{u}_x}{\uD\/t}\ =\ \gamma\,\frac{y+d}{h}, 
\end{align}
where $\gamma$ is the vertical acceleration at the free surface 
\begin{align}\label{defgamma}
  \gamma\ \equiv\ \left.\frac{\uD\,v}{\uD\/t}\right|_{y=\eta}\ \approx\ h\left[\,\bar{u}_x^{\,2}\,-\,\bar{u}_{xt}\,-\,\bar{u}\,\bar{u}_{xx}\,\right].
\end{align}

The kinetic $\mathscr{K}$ and the potential energies of gravity $\mathscr{V}_g$ and capillarity $\mathscr{V}_c$ are 
\begin{align}
  \frac{\mathscr{K}}{\rho}\ &\equiv\ \int_{t_1}^{t_2} \int_{x_1}^{x_2}\int_{-d}^\eta\frac{u^2+v^2}{2}\,\ud\/y\,\ud\/x\,\ud\/t\ \approx\ \int_{t_1}^{t_2} \int_{x_1}^{x_2}\left[\,\frac{h\,\bar{u}^2}{2}\ +\ \frac{h^3\,\bar{u}_x^{\,2}}{6}\,\right]\ud\/x\,\ud\/t, \label{defK} \\
  \frac{\mathscr{V}_g}{\rho}\ &\equiv\ \int_{t_1}^{t_2} \int_{x_1}^{x_2}\int_{-d}^\eta g\,(y+d)\,\ud\/y\,\ud\/x\,\ud\/t\ =\ \int_{t_1}^{t_2} \int_{x_1}^{x_2}\frac{g\,h^2}{2}\,\ud\/x\,\ud\/t, \label{defVg} \\
  \frac{\mathscr{V}_c}{\rho}\ &\equiv\ \int_{t_1}^{t_2} \int_{x_1}^{x_2}\tau\left(\,\sqrt{1+h_x^{\,2}}\,-\,1\right) \ud\/x\,\ud\/t, \label{defVc} 
\end{align}
where $\rho$ is the (constant) density of the fluid, $g$ is the (constant) acceleration due to gravity (directed downward) and $\tau$ is a (constant) surface tension coefficient (divided by the density). An action integral $\mathscr{S}$ (temporal integral of a Lagrangian) can then be introduced as the kinetic minus the potential energies plus a constraint for the mass conservation (Hamilton principle), \ie
\begin{equation} \label{defS}
  \frac{\mathscr{S}}{\rho}\ \equiv\ \int_{t_1}^{t_2} \int_{x_1}^{x_2}\left[\,\frac{h\,\bar{u}^2}{2}\ +\ \frac{h^3\,\bar{u}_x^{\,2}}{6}\ -\ \frac{g\,h^2}{2}\ +\ \tau\ -\ \tau\,\sqrt{1+h_x^{\,2}}\ +\,\left\{\,h_t\,+\left[\,h\,\bar{u}\,\right]_x\,\right\}\phi\,\right]\ud\/x\,\ud\/t, 
\end{equation}
where $\phi$ is a Lagrange multiplier for the term enforcing the mass conservation.

The Euler--Lagrange equations for the functional \eqref{defS} are
\begin{align}
  \delta\phi:\ & 0\ =\ h_t\ +\,\left[\,h\,\bar{u}\,\right]_x,\label{dLdphi}\\
  \delta\bar{u}:\ & 0\ =\ \phi\,h_x\ -\ [\,h\,\phi\,]_x\ -\ \third\,[\,h^3\,\bar{u}_x\,]_x\ +\ h\,\bar{u}, \\
  \delta h:\ & 0\ =\ \half\,\bar{u}^2\ +\ \half\,h^2\,\bar{u}_x^{\,2}\ -\ g\,h\ +\ \tau\left[\,h_x\left(1+h_x^{\,2}\right)^{-1/2}\,\right]_x\ -\ \phi_t\ +\ \phi\,\bar{u}_x\ -\ [\,\bar{u}\,\phi\,]_x,
\end{align}
thence
\begin{align}
  \phi_x\ &=\ \bar{u}\ -\ \third\,h^{-1}\,[\,h^3\,\bar{u}_x\,]_x,  \label{serphix}\\\phi_t\ &=\ \half\,h^2\,\bar{u}_x^{\,2}\ -\ \half\,\bar{u}^2\ -\ g\,h\ +\ \tau\left[\,h_x\left(1+h_x^{\,2}\right)^{-1/2}\,\right]_x +\ \third\,\bar{u}\,h^{-1}\,[\,h^3\,\bar{u}_x\,]_x. \label{serphit}
\end{align}
The variable $\phi$ can be easily eliminated from the equations \eqref{serphix} and \eqref{serphit}, and several secondary equations can be obtained subsquently. Thus, after some algebra, one gets  
\begin{align} 
  h_t\ +\,\left[\,h\,\bar{u}\,\right]_x\ &=\ 0,  \label{eqmasse} \\
  \left[\,\bar{u}\,-\,\frac{(h^3\/\bar{u}_x)_x}{3\,h}\,\right]_t\ +\,\left[\,\frac{\bar{u}^2}{2}\,+\,g\,h\,-\,\frac{h^2\,\bar{u}_x^{\,2}}{2}\,-\,\frac{\bar{u}\,(h^3\/\bar{u}_x)_x}{3\,h}\,-\,\frac{\tau\,h_{xx}}{\left(1+h_x^{\,2}\right)^{3/2}}\,\right]_x\ &=\ 0, \label{eqqdmbisse} \\
  \left[\,h\/\bar{u}\,-\,\frac{(h^3\bar{u}_x)_x}{3}\,\right]_t\ +\,\left[\,h\/\bar{u}^2\,+\,\frac{g\/h^2}{2}\,-\,\frac{2\/h^3\/\bar{u}_x^{\,2}}{3}\,-\,\frac{h^3\/\bar{u}\/\bar{u}_{xx}}{3}\,-\,h^2\/h_x\/\bar{u}\/\bar{u}_x\,-\,\tau\/R\,\right]_x\, &=\ 0, \label{eqqdmfluxbisse} \\
  \left[\,h\,\bar{u}\,\right]_t\, +\,\left[\,h\,\bar{u}^2\,+\,\half\,{g\,h^2}\,+\,\third\,{h^2\,\gamma}\,-\,\tau\,R\,\right]_x\ &=\ 0, \label{eqqdmfluxse} \\
  \left[\,\frac{h\,\bar{u}^2}{2}\,+\,\frac{h^3\,\bar{u}_x^{\,2}}{6}\,+\,\frac{g\,h^2}{2}\,+\,\tau\/\sqrt{1+h_x^{\,2}}\,\right]_t\ +\ &\nonumber\\
  \left[\left(\frac{\bar{u}^2}{2}\,+\,\frac{h^2\,\bar{u}_x^{\,2}}{6}\,+\,g\,h\,+\,\frac{h\,\gamma}{3}\,\,-\,\frac{\tau\,h_{xx}}{\left(1+h_x^{\,2}\right)^{3/2}}\,\right)h\,\bar{u}\,+\,\frac{\tau\,h_{x}\,(h\bar{u})_x}{\left(1+h_x^{\,2}\right)^{1/2}}\,\right]_x\, &=\ 0. \label{eqenese}
\end{align}
where
\begin{equation*}
  R\ \equiv\ h\,h_{xx}\left(1+h_x^{\,2}\right)^{-3/2}\ +\,\left(1+h_x^{\,2}\right)^{-1/2}.
\end{equation*}
Physically, these equations characterise the conservations: of the mass \eqref{eqmasse}, of the tangential momentum at the free surface \eqref{eqqdmbisse}, of the momentum flux \eqref{eqqdmfluxbisse}--\eqref{eqqdmfluxse} and of the energy \eqref{eqenese}.

These equations, without surface tension, were first derived by \textsc{Serre} \cite{Serre1953a}, independently rediscovered by Su and Gardner \cite{SG1969}, and again by \textsc{Green}, \textsc{Laws} and \textsc{Naghdi} \cite{Green1974}. These approximations are valid in shallow water without assuming small amplitude waves, so they are sometimes called {\em weakly-dispersive fully-nonlinear approximation} \cite{Wu2001a} and are a generalisation of the \textsc{Saint--Venant} and of the \textsc{Boussinesq} equations.


\section{Steady waves}
\label{sec:steady}

The previous equations being Galilean invariant, we consider now a steady wave motion (\ie a frame of reference moving with the wave where solutions are independent of time). For $2L$-periodic solutions, the mean water depth $d$ and the mean depth-averaged velocity $-c$ are
\begin{equation} \label{defdc}
  d\ =\,\left<\,h\,\right>\,=\ \frac{1}{2L}\int_{-L}^{L}h\,\ud\/x, \qquad -\,c\,d\ =\,\left<\,h\,\bar{u}\,\right>\, =\ \frac{1}{2L}\int_{-L}^{L}h\,\bar{u}\,\ud\/x, 
\end{equation}
thus $c$ is the wave phase velocity observed in the frame of reference without mean flow. The mass conservation \eqref{eqmasse} yields
\begin{equation} \label{relcuste}
  \bar{u}\ =\ -\,c\,d\left/\,h\right.
\end{equation}

Let $\Fr=c^2/gd$ be the Froude number squared and let $\Bo=\tau/gd^2$ be a Bond number; for later convenience, we introduce also the Webber number $\We=\Bo/\Fr=\tau/c^2d$. Substitutions of \eqref{relcuste} into \eqref{eqqdmbisse} and \eqref{eqqdmfluxse}, followed by one integration, give 
\begin{align}
\frac{\Fr\,d}{h}\ +\ \frac{h^2}{2\,d^2}\ +\ \frac{\gamma\,h^2}{3\,g\,d^2}\ -\ 
\frac{\Bo\,h\,h_{xx}}{\left(1+h_x^{\,2}\right)^{3\over2}}\ -\ 
\frac{\Bo}{\left(1+h_x^{\,2}\right)^{1\over2}}\ 
=\ \Fr\ +\ \frac{1}{2}\ -\ \Bo\ +\ \mathscr{K}_1, \label{eqqdmfluxseperm} \\
\frac{\Fr\,d^2}{2\,h^2}\ +\ \frac{h}{d}\ +\ \frac{\Fr\,d^2\,h_{xx}}{3\,h}\ 
-\ \frac{\Fr\,d^2\,h_{x}^{\,2}}{6\,h^2}
-\ \frac{\Bo\,d\,h_{xx}}{\left(1+h_x^{\,2}\right)^{3\over2}}\ 
=\ \frac{\Fr}{2}\ +\ 1\ +\ \frac{\Fr\,\mathscr{K}_2}{2}, \label{eqqdmbisseperm}
\end{align}
with
\begin{equation*}
  {\gamma}\,/\,{g}\ =\  {\Fr\,d^3\,h_{xx}}\,/\,{h^2}\ -\ {\Fr\,d^3\,h_{x}^{\,2}}\,/\,{h^3},
\end{equation*}
$\mathscr{K}_n$ being dimensionless integration constants to be determined exploiting the conditions \eqref{defdc} ($\mathscr{K}_1 = \mathscr{K}_2 = 0$ for solitary waves). Computing \eqref{eqqdmfluxseperm}$-(h/d)\times$\eqref{eqqdmbisseperm} in order to eliminate $h_{xx}$ between \eqref{eqqdmfluxseperm} and \eqref{eqqdmbisseperm}, one obtains easily
\begin{align}
  \frac{\Fr\/d}{2\/h}\, -\, \frac{h^2}{2\/d^2}\, -\, \frac{\Fr\/d\/h_{x}^{\,2}}{6\/h}\, -\, \frac{\Bo}{\left(1+h_x^{\,2}\right)^{1\over2}}\, =\, \Fr\, +\, \frac{1}{2}\, -\, \Bo\, +\, \mathscr{K}_1\, -\, \frac{(\Fr\/+\/2\/+\/\Fr\/\mathscr{K}_2)\/h}{2\/d}, \label{ode1h}
\end{align}
that is a first-order ordinary differential equation for $h$.

The integration constants $\mathscr{K}_n$ are defined from the conditions \eqref{defdc}. Averaging the equations \eqref{eqqdmbisseperm} and $(d/h)\times$\eqref{eqqdmfluxseperm} yields
\begin{align}
  \mathscr{K}_2\ &=\, \left<\,\frac{(3+h_{x}^{\,2})\,d^2}{3\,h^2}\,-\,1\,\right>, \label{defK2}\\
  \mathscr{K}_1\,+\,\half\,+\,\Fr\,-\,\Bo\ &=\,\left. 
  \left<\,\frac{\Fr\,d^2}{h^2}\,+\,\frac{1}{2}\,-\, 
  \frac{\Bo\,d\,/\,h}{\left(1+h_x^{\,2}\right)^{1\over2}}\,\right>\right/
  \left<\,\frac{d}{h}\,\right>. 
\end{align}
For infinitesimal periodic waves, we have $h\approx d+a\cos(kx)$ with $|a/d|\ll1$, thence $\mathscr{K}_1 \approx \mathscr{K}_2\approx0$ and the linearised dispersion relation $\Fr \approx \left[1 + \Bo(kd)^2\right]\left/\left[1 + (kd)^2/3\right]\right.$ is obtained.


\subsection{Solitary waves}

Considering solitary waves --- \ie $h(\infty)=d$ thence $\mathscr{K}_1 = \mathscr{K}_2 = 0$ --- and writing $h'=\ud h(x)/\ud x$, the equation \eqref{ode1h} multiplied by $(-\/2\/h\left/\/d\right.)$ becomes
\begin{align}\label{ode1F}
  F(h',h)\, \equiv\,\frac{\Fr\/{h'}^2}{3}\, +\, \frac{2\/\Bo\/h/d}{\left(1 + {h'}^2\right)^{1\over2}}\, -\, \Fr\, +\, \frac{(2\Fr+1-2\Bo)\/h}{d}\, -\, \frac{(\Fr+2)\/h^2}{d^2}\, +\, \frac{h^3}{d^3}\, =\, 0,
\end{align}
with the partial derivatives
\begin{align}
  F_h\ &=\ \frac{2\,\Bo}{d}\left(1+{h'}^2\right)^{-{1\over2}}\ +\ \frac{(2\Fr+1-2\Bo)}{d}\ -\ \frac{2\,(\Fr+2)\,h}{d^2}\ +\ \frac{3\,h^2}{d^3}, \label{ode1Fh} \\
  F_{h'}\ &=\ \frac{2\,\Fr\,{h'}}{3}\ -\ 2\,\Bo\,h'\,\frac{h}{d}\left(1 + {h'}^2\right)^{-{3\over2}}. \label{ode1Fhp}
\end{align}
These derivatives always exist and are bounded. However, we can have $F_{h'}=0$, so singular points may exist. They are investigated below.


\subsection{Limiting cases for solitary waves}

We consider here several limiting cases of equation \eqref{ode1F} which deserve a particular attention.

\subsubsection{Pure gravity waves}

For pure gravity waves $\Bo=0$, letting $\Fr\ =\ 1\ +\ a\,/\,d$, the solitary wave solution is 
\begin{equation*}
  h\ =\ d\ +\ a\operatorname{sech}^2(\kappa x/2),\qquad (\kappa\/d)^2\ =\ 3\,a\,/\,(d+a).
\end{equation*}
that is explicit and well-known since the work of Serre \cite{Serre1953}. Note that $a$ should be nonnegative.

\subsubsection{Pure capillary waves}

Pure capillary waves are obtained when $g=0$ that is letting $\Fr\to\infty$ and $\Bo\to\infty$ but keeping $\We$ constant. The equation \eqref{ode1F} divided by $\Fr$ then becomes
\begin{align}
  F(h',h) \, \equiv\, \frac{{h'}^2}{3}\ +\ \frac{2\,\We\,h/d}{\left(1 + {h'}^2\right)^{1\over2}}\ -\ 1\ +\ \frac{2\,(1-\We)\,h}{d}\ -\ \frac{h^2}{d^2}\ =\ 0. \label{ode1Fcap}
\end{align}
This equation is still cubic in $(1+h'^2)^{1\over2}$ but only quadratic in $h$.

Clearly, $F$ vanishes at the point $h=d$, $h'=0$. The Taylor expansion near $h=d, h'=0$ at order $4$ (\ie near $x=\pm\infty$) gives
\begin{equation}
  F(h',h)\ \approx\ (1/3-\We)\,h'^2\ -\ (h-d)^2\ -\ \We\,(h-d)\,h'^2\ +\ (3/4)\,\We\,h'^4,
\end{equation}
that is a quadratic equation in  ${h'}^2$. So for $\We > 1/3$, the only solution starting at $(h=d, h'=0)$ is the constant solution $h = d$. For values $\We \leqslant 1/3$, since $F(0,h) = -(\frac{h}{d}-1)^2$, as we shall see below, there are no regular solitary waves. In the next section we shall introduce singular solution which correspond to angular solitary waves (``\emph{peakons}'') with zero semi-curvatures. However it can be shown that for pure capillary equation, there are no such singular solutions.

\subsubsection{Approximation for small slopes}

Assuming that $|h'| \ll 1$, one can reasonably use the approximation
\begin{align*}
  \left(1 + {h'}^2\right)^{-{1\over2}}\ \approx\ 1\ -\ \half\,{h'}^2, 
\end{align*}
and the equation \eqref{ode1F} becomes
\begin{align}
  \left(\frac{\Fr}{3}-\frac{\Bo\,h}{d}\right)h'^2\ \approx\ \Fr\ -\ \frac{(2\Fr+1)\,h}{d}\ +\ \frac{(\Fr+2)\,h^2}{d^2}\ -\ \frac{h^3}{d^3}. \label{ode1Fapp}
\end{align}
The solutions of this equation can be obtained analytically, as for pure gravity waves. Indeed, with the change of independent variable
\begin{equation*}
  \ud\/\xi\ =\ \left|\,1\,-\,3\,\We\,h\,/\,d\,\right|^{-{1\over2}}\,\ud\/x,
\end{equation*}
the equation \eqref{ode1Fapp} yields
\begin{align*}
  \frac{\Fr}{3}\left(\frac{\ud\,h}{\ud\/\xi}\right)^{\!2}\ =\ \Fr\ -\ \frac{(2\Fr+1)\,h}{d}\ +\ \frac{(\Fr+2)\,h^2}{d^2}\ -\ \frac{h^3}{d^3}, 
\end{align*}
that can be explicitly solved
\begin{align*}
  h(\xi)\ &=\ d\ +\ a\operatorname{sech}^2(\kappa\xi/2),\qquad (\kappa\/d)^2\ =\ 3\,a\,/\,(d+a), \qquad \Fr\ =\ 1\ +\ a\,/\,d,
\end{align*}
and $x(\xi)$ can be also obtained explicitly as a complicated expression (not given here as it is secondary for the purpose of the present paper).


\subsection{Regular waves}

For regular waves, the crest at $x = 0$ is smooth with $h'(0) = 0$ and $h(0) = d+a$, $a$ being the wave amplitude (not necessarily positive). The equation \eqref{ode1F} then yields
\begin{equation*}
  \Fr\ =\ 1\ +\ a\,/\,d,
\end{equation*}
thence the amplitude is independent of the Bond number $\Bo$.


\subsection{Asymptotic analysis}
\label{secAsymp}

For solitary waves decaying exponentially in the far field, we have $h\sim d+a\exp(-\kappa x)$ as $x\to+\infty$, where $\kappa$ is a trend parameter such as $\Re(\kappa)>0$. The equation \eqref{ode1F} then yields  the relation
\begin{equation}\label{disprel}
  \Fr\ =\ \frac{3\,-\,3\,\Bo\,(\kappa\/d)^2}{3\,-\,(\kappa\/d)^2} \qquad \text{or}\qquad (\kappa\/d)^2\left(\Fr\,-\,3\,\Bo\right)\, =\ 3\left(\Fr\,-\,1\right).
\end{equation}
Since $\Fr$ and $\Bo$ are both real numbers, this relation shows that $\kappa$ should be either real or pure imaginary, \ie there are no solitary waves with damped oscillations (and with exponential decay). The special case $\Bo = 1/3$ yields $\Fr = 1$ or $\kappa d = \sqrt{3}$. Higher-order terms (not given here) then show that $\Fr = 1$ and $\kappa d=\sqrt{3}$. Conversely, the special case $\Fr = 1$ yields $\Bo = 1/3$ and $\kappa d$ is undefined from the relation \eqref{disprel}, but higher terms give $\kappa d = \sqrt{3}$. Possible solutions of this type are investigated in the section \ref{secsolexp} below.

For solitary waves decaying algebraically in the far field, we have $h\sim d\/ + \/a(\kappa\/x)^{-\alpha}$ as $x\to+\infty$, where $\alpha>1$ is a parameter. The equation \eqref{ode1h} then yields, necessarily, that $\Fr=1$. Thus, if algebraic solution exist, they must occur at the critical Froude number. Considering higher-order terms (not given here), integer values of $\alpha > 2$ suggest that we should have $\Bo = 1/3$. For $\alpha=2$, similar consideration suggest that there may be algebraic solitary waves when $\Bo \neq 1/3$.

The asymptotic analysis provides only partial information. The techniques of the phase space analysis, exposed in the section \ref{secsolexp} below, allow to obtain a more complete information on the qualitative behaviour of solutions and unveil new interesting phenomena. In particular, we shall see that algebraic solitary waves exist only as weak solutions, \ie with an angular crest.


\subsection{Singular values}

Singular points are such that $F_{h'}=0$. This condition leads to three real possibilities, denoting $h_0=h(0)$ and $h'_0=h'(0)$,
\begin{equation}
  h'\ =\ h_0'\ \equiv\ 0, \qquad h'\ =\ h_\pm'\ \equiv\ \pm\left[\left(\frac{\,3\,\Bo\,h}{\Fr\,d}\right)^{\!{2\over3}}-1\,\right]^{1\over2},
\end{equation}
where the condition $h/d \geqslant \Fr/3\Bo$. The case $h'=0$ yields, from $F = 0$, that $h = d$ or $h = \Fr\/d$.


\section{Phase plane analysis for solitary waves}
\label{secsolexp}

In this section, we apply algebraic techniques to the phase plane analysis of solitary waves. In coherence with the asymptotic analysis, we assume that when $x$ tends to plus or minus infinity
\begin{equation}\label{eq:infty}
  h(\infty)\ =\ d, \qquad  h'(\infty)\ =\ 0.
\end{equation}
Note that $h=d$ is therefore a trivial solution of our problem. Since in equation \eqref{ode1F}, $h'$ only appears by its square $h'^2$ and $h(-\infty) = h(\infty)$, if $h(x)$ is a solution of the equation then $h(-x)$ is also a solution. Therefore, our equation admits symmetric solutions, but the whole set of solutions is not limited to symmetric ones.

For the convenience of the reader and the clarity of the pictures, we make the following mild simplifications: we focus on the cases where $(\Fr, \Bo) \in [0,2]\times[-1,2]$ (in accordance with physical reasons) and, when simplification is needed, we normalise $d$ to $1$ without loss of generality. We use subscripts to indicate the parameters $(\Fr,\Bo)$ and omit them when there are no ambiguities.

To explain our approach, let us notice that our generalisation of SGN equation takes into account gravity, inertia and capillarity. The balance between the effects of the formers is expressed by the parameter $\Fr$: when $\Fr>1$, the kinematic energy dominates the potential energy, correlatively the solitary waves will be of elevation; while when $\Fr<1$ the solitary wave is of depression. Similarly, the balance between the effects of capillarity and gravity (resp. inertia) is expressed by the parameter $\Bo$ (resp. $\We$), but now the transition appear when $\Bo=1/3$ (respectively $\We=1/3$). To these different effects (hence behaviours of the solutions) is associated a first rough partition of the parametric plane $(\Fr, \Bo)$, delimited by the three lines $\Fr = 1$, $\Bo =1/3$ and  $\Fr=3\Bo$. We aim to refine this partition, by considering additional curves in the parametric plane, in order to classify all the possible appearances of solitary waves.

In order to discuss the number and behaviour of the solutions of the differential equation $F_{\Fr,\Bo}(h',h) = 0$, with respect to the pair of parameters $(\Fr,\Bo) \in [0,2]\times[-1,2]$, our approach is to describe graphically the variations of $h$ and $h'$, through the corresponding family of real algebraic curves $C_{\Fr,\Bo} \in \R^2$. The implicit equation of $C_{\Fr,\Bo}$ is $F_{\Fr,\Bo}(h',h)=0$. We will decompose the parameters space $(\Fr,\Bo)$, with our restriction to the rectangle $[0,2]\times[-1,2]$, into subdomains where the plane curves $C_{\Fr,\Bo}$ have the same ``shape'' (in particular the same topology). Some subdomains can be very small, so to overcome the resulting computational difficulties, the study of the curves delimiting the subdomains will be performed with a certified topology methods (see, \eg \cite{Gonzalez-Vega2002,Cheng2010} and the references there in).


\subsection{An example of phase diagram}
\label{sec:Ex1}

\begin{figure}
  \centering
  \subfigure[]{\includegraphics[width=0.48\textwidth]{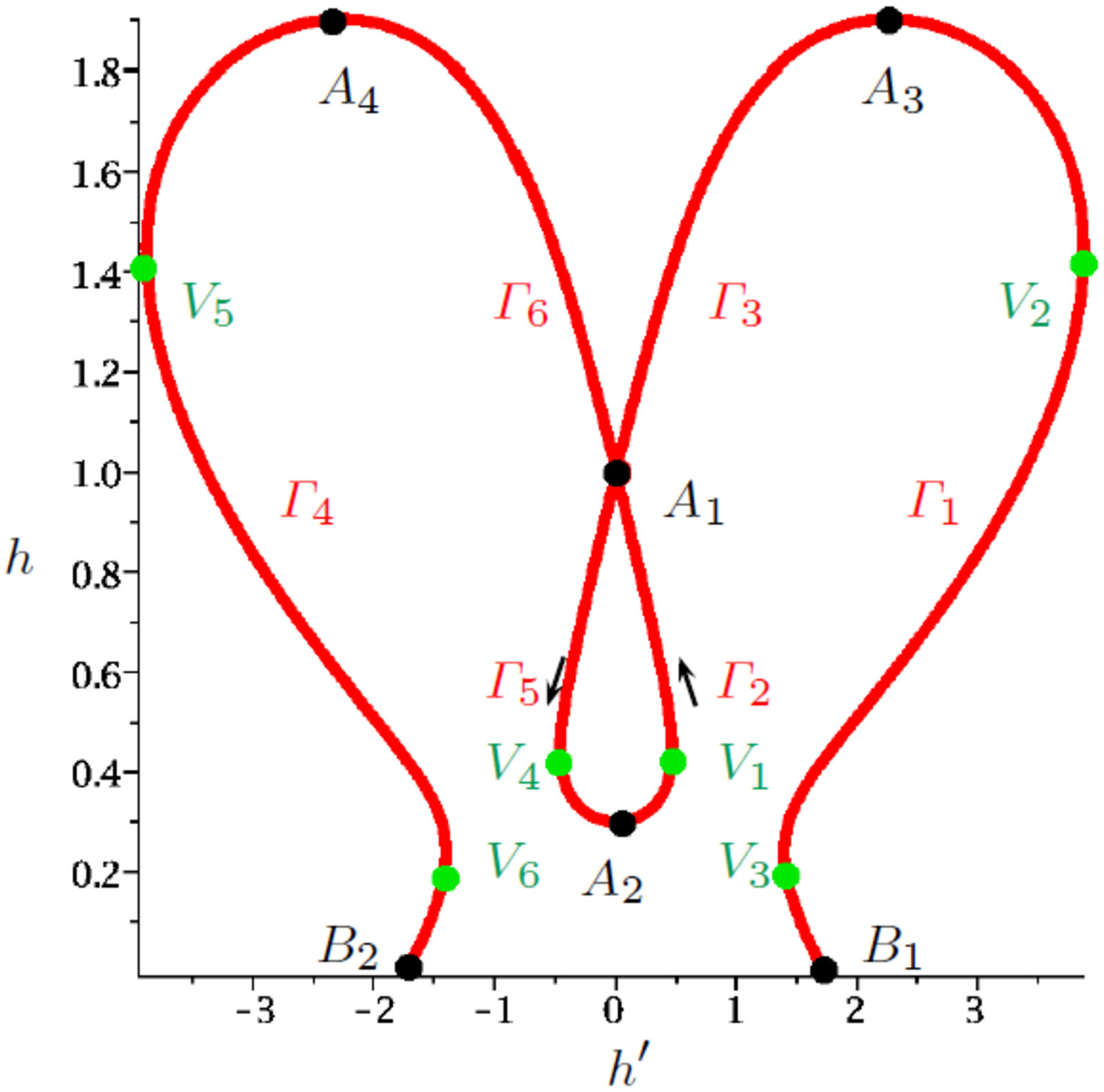}}
  \subfigure[]{\includegraphics[width=0.48\textwidth]{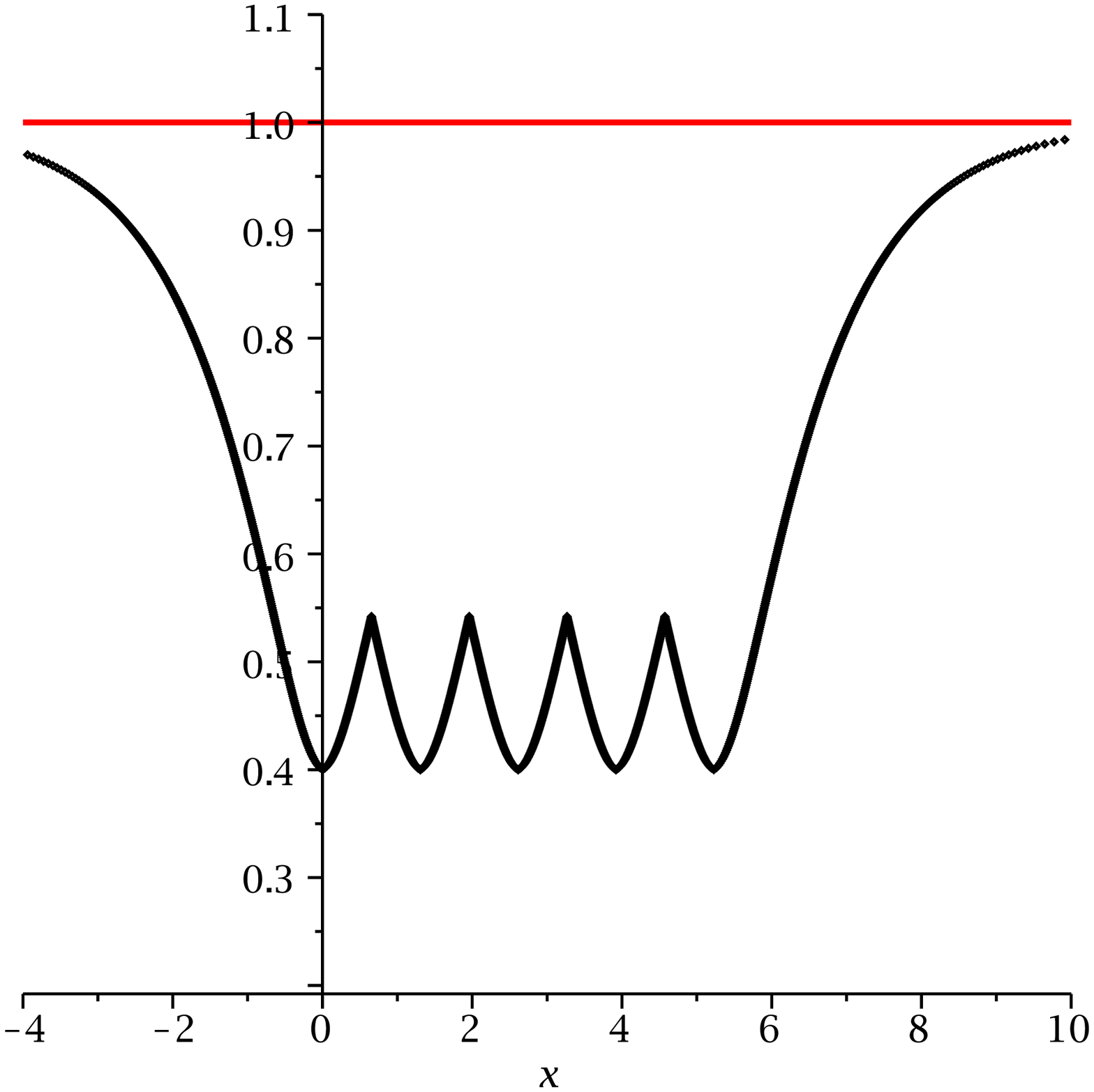}}
  \caption{\small\em (a) Phase space diagram from Section~\ref{sec:Ex1}. (b) Multiangular wave.}
  \label{fig:ex1}
\end{figure}

Let us illustrate our approach with the phase plane description for some value of the parameters, for instance, $\Fr=0.3<1$ and $\Bo=0.8>1/3$. The corresponding curve $C_{0.3,0.8}$ is shown in the left panel of Figure~\ref{fig:ex1}. Since the values $0.3$ and $0.8$ are exact rational numbers, it can be plotted simply relying on a discretisation, \eg with \textsc{Matlab} or drawn with a certified topology technique using specialised programs such as the algebraic curves package of \textsc{Maple} or \textsc{Axel} (see \url{http://axel.inria.fr}). In particular, this means that loops of whatever size are preliminary detected relying on exact (given certain precision) numerical and algebraic computations. Note that the curve is symmetric with respect to the $h$-axis. This geometric property is general in our model, since the equations involve only the square of the derivative $h'$.

After computing the points of the curve with a horizontal tangent and the singular points, we can decompose the curve into a finite number $N$ of graphs portions $\Gamma_j$, $j=1,\cdots,N$ of type $h'=\phi_j(h)$, for some differentiable functions $\phi_j$, $j = 1,\cdots,N$ according to the implicit function theorem. The end points of each graph correspond to critical points of the (horizontal) projection of the curve on the $h$ axis. The points with a horizontal tangent or multiple points satisfy the additional explicit condition $\partial_{h'}F = 0$. On every graph, we have an explicit (possibly nonlinear) ODE $\ud x=\ud h/\phi_j(h)$ which can be integrated by standard techniques. Moreover, where $h'$ is positive (respectively negative), $h$ should increase (resp. decrease), so each $\Gamma_j$ can be oriented (see Figure~\ref{fig:ex1} where the curve can be decomposed into $N = 6$ oriented curve segments).

In this particular example, there are four points on the curve such that $\partial F/\partial h' = 0$: $A_1=(0,1)$, $A_2=(0,0.3)$, $A_3\approx(2,1.9)$ and $A_4\approx(-2,1.9)$. Call $B_1$ and $B_2$ the intersections of the curve with the $h'$ axis. Then, we get the following going up branches: $\Gamma_1$ from $B_1$ to $A_3$, $\Gamma_2$ from  $A_2$ to $A_1$ (on the right side) and $\Gamma_3$ from $A_1$ to $A_3$. Symmetrically, the down-going branches $\Gamma_j$ for $j=\{4,5,6\}$ are in the left half-plane $h' < 0$.

Constrained by the asymptotic conditions \eqref{eq:infty}, we only consider homoclinic paths starting from $A_1$ and arriving at $A_1$. Topologically they correspond to loops. In this example, the only continuous such loop is going down on $\Gamma_5$ from $A_1$ to $A_2$ (on the left side), then going up on $\Gamma_2$ from $A_2$ to $A_1$ (on the right side). It will correspond to the unique differentiable solution $h(x)$ of the differential equation.

Inspired by the examples computed by Stokes \cite{Stokes1880} and by the analysis of the ``peakons'' of some nonlinear dispersive equations such as the Camassa--Holm, Degasperis--Procesi and other equations \cite{Camassa1993, Degasperis1999, Liao2013, Liao2014}, we also consider a different class of solutions, which are sometimes referred to as {\em singular solutions\/} in the classical literature \cite{Arnold1996, Hamburger1893, Hubert1996}. Here, they correspond to solitary waves with one (or several) angular points. However, since there are an infinity of angular solutions, we additionally require that the two semi-derivatives of $h(x)$ at the angle are opposite in sign and that the two second semi-derivatives vanish. This property was put in evidence by Stokes in his limiting wave construction. Physically, this means that the wave is symmetric at the crest and that the fluid vertical acceleration is zero at the crest.

In total, there are six points on the curve $C_{0.3,0.8}$ with a vertical tangent line. They have approximately the following coordinates: $V_1 = (0.5, 0.4)$, $V_2 = (4, 1.4)$, $V_3 = (1.3, 0.2)$ and their symmetric counterparts, $V_4$, $V_5$ and $V_6$, with respect to the $h$-axis. The two semi second derivatives at these points are zero. On the one hand, two points $V_2$ and $V_3$, and their symmetric counterparts, belong to graphs not connected to $A_1$. On the other hand, $V_1$ and $V_4$ are connected to $A_1$ respectively by the $\Gamma_2$ and $\Gamma_5$ branches. So, we can form a discontinuous symmetrical loop going down from $A_1$ to $V_4$ (on $\Gamma_5$), jumping from $V_4$ to $V_1$, then going up from $V_1$ to $A_1$ (on $\Gamma_1$). It corresponds to a new solitary wave with an angular point, a so-called {\em peakon}, $h(x)$ satisfying  all the conservation laws. We can consider as well another loop going down from $A_1$ to $A_2$ (on the left side), then going up from $A_2$ to $V_1$ (on the right side), then jumping from $V_1$ to $V_4$, and cycling through $V_4, A_2, V_1, V_4$, then finally going up from $A_2$ to the boundary point $A_1$. The corresponding solution has a finite number of `peaks' (see Figure~\ref{fig:ex1}(\textit{b})). The angular points are emphasised in Figure~\ref{fig:ex1}(\textit{b}) by green solid disks.

Additionally, one can construct a discontinuous path going from $A_1$ to $A_3$, jumping to $A_4$ and going back to $A_1$. This will correspond to a singular solitary wave with a cusp at the crest. However, such solutions should be discarded based on physical considerations, since it leads to infinite accelerations in the singularity point.

We now aim to generalise the analysis given above for $\Fr=0.3$ and $\Bo=0.8$ to all pairs $(\Fr,\Bo) \in [0,2]\times[-1,2]$. Clearly, for a very small perturbation $\Fr=0.3+\epsilon_1$ and $\Bo=0.8+\epsilon_2$, the curves remain qualitatively very similar. In fact, one can pass from the curve $C_{0.3, 0.8}$ to $C_{0.3+\epsilon_1, 0.8+\epsilon_2}$ by a homotopy respecting the decomposition in oriented graphs we have described. However, for larger perturbations of $\Fr$ and $\Bo$ numbers, we may have qualitatively different curves and results. For instance, under a deformation, the considered loop retracts to one point and disappears completely; so do solitary wave solutions described above. This abrupt change of behaviour is usually called a bifurcation. In our application, the bifurcations may occur when one of the numbers of special points (multiple point, point with a vertical or a horizontal tangent) of the curve $C_{\Fr,\Bo}$ changes.

Our purpose is to describe all possible situations (up to qualitative similarity) and to delimit the corresponding parameters $(\Fr,\Bo)$ ranges. Due to the algebraic nature of the problem, these regions are organised in a finite number of semi-algebraic sets, \ie the equations of their borders are polynomials in $(\Fr,\Bo)$ that can be effectively computed. On the borders of the open domains, we will also have special extremal behaviours, which are described below.


\subsection{Local analysis where $h' = 0$}

The points on the $h$-axis are important because they correspond to the wave crest or trough. To determine them, it is sufficient to substitute $h'=0$ and solve the resulting equation in $h$ $F_{\Fr,\Bo}(0,h)=0$, that gives either $h=d$ with multiplicity $2$, or $h=d\Fr$. If there is a branch of type $\Gamma$ connecting these two points, then there should be also the symmetric one. Consequently, we can consider the path starting from $A_1 = (h' = 0,\, h = d)$ and going up (respectively down) to $A_2=(h'=0, h=d\Fr)$ if $\Fr>1$ (respectively $\Fr<1$); and then back from $A_2$ to $A_1$. This could correspond to the crest (trough) of a solitary wave, above (below) the still water level $h = d$.

Then, in order to investigate the local behaviour at $A_1=(h'=0,h=d)$, we compute the Taylor expansion of $F$ at this point. It is similar to the asymptotic analysis conducted in Section~\ref{secAsymp}. By expanding the equation to the second order, we obtain
\begin{equation*}
  d^2\,(\Fr-3\Bo)\,{h'}^2\ -\ 3\,(\Fr-1)\,(h-d)^2\ =\ 0\ + \ \mathrm{O}\!\left((h-d)^3,{h'}^4\right).
\end{equation*}

There are three distinct cases to consider depending on the sign of the expression $(\Fr-1)(\Fr-3\Bo)$:
\begin{itemize}
  \item If $(\Fr-1)(\Fr-3\Bo)<0$, $A_1$ is an isolated point of the curve $C_{\Fr,\Bo}$. This implies that the only solution is the trivial one $h = d$.
  \item If $(\Fr-1)(\Fr-3\Bo)>0$, $A_1$ is a double point corresponding to the crossing of two branches of $C_{\Fr,\Bo}$. Hence, there is no obstruction (at least at the level of this local analysis) for the existence of a solitary wave.
  \item If $(\Fr-1)(\Fr-3\Bo)=0$, with $\Fr=1$. Then $A_1=A_2$ and we must consider the Taylor expansion up to the third-order
  \begin{equation}\label{eq:cusp}
    d^3\,(1-3\Bo)\,{h'}^2\ +\ 3\,(h-d)^3\ -\ d^2\,(h-d)\,{h'}^2\ =\ 0.
  \end{equation}
  Therefore, if $\Bo>1/3$ (respectively $\Bo<1/3$), the curve $C_{\Fr,\Bo}$ admits at $A_1$ a cusp above (resp. below) $A_1$. While in the case $\Bo = 1/3$, the Taylor expansion corresponds to three smooth branches including $h = d$. As a consequence, there is no obstruction (in this local analysis) for the existence of a solitary wave solution with an angular point (\ie a peakon).
  \item If $(\Fr-1)(\Fr-3\Bo) = 0$, with $\Fr = 3\Bo$ and $\Fr \neq 1$. In this case, the Taylor expansion to the third order is
  \begin{equation*}
    (h-d)\,[ (3\Bo -1)\,(h-d) - \Bo\ d^2\, {h'}^2\  =\ 0,
  \end{equation*} 
  that gives locally a line ($h=d$) and a smooth curve (a parabola). As a consequence, there is no obstruction for the existence of a non trivial solitary wave solution with an angular point and an algebraic decrease at infinity.
\end{itemize}

Now, we perform a similar local analysis for the point $A_2$. The Taylor expansion to the second order of the equation $F(h', h) = 0$ at the point $A_2 = (h'=0,h=d\Fr)$, yields
\begin{equation*}
  3\,(\Fr-1)^2\,(h-d\Fr)\ =\ d\,\Fr\left(3\Bo-1\right){h'}^2.
\end{equation*}
Thus, when $\Fr\neq 1$ and $\Bo\neq1/3$, we obtain a regular point with a horizontal tangent. It is of convex type if $\Bo>1/3$ and concave if $\Bo<1/3$. Therefore, if $\Fr<1$ and $\Bo>1/3$, or if $\Fr>1$ and $\Bo<1/3$, there is no obstruction for the existence of a solitary wave. While if $\Fr<3\Bo<1$ or if $\Fr>3\Bo>1$, this local analysis allows only an angular solitary wave.

When $\Fr\neq1$ and $\Bo=1/3$, the Taylor expansion of the equation at the point $A_2$ yields
\begin{equation*}
  \Fr {h'}^4\ \sim \, 4 (\Fr -1)^2 (h-d).
\end{equation*}
In other words, the curve has a flat point (the curvature vanishes).

As a main conclusion of this local analysis, the parameter plane $(\Fr,\Bo)$ is partitioned into six regions delimited by the three straight lines $\Fr=1$, $\Bo=1/3$ and $\Fr=3\Bo$. We would like to underline that these lines correspond to critical physical regimes.


\subsection{Local analysis where $h'\neq 0$}

As noted in the end of Section~\ref{sec:Ex1}, a solitary wave is associated to a path on the curve $C_{\Fr,\Bo}$ from the point $A_1=(0,1)$ to a point with a horizontal or vertical tangent. Therefore, these points play a key role in our analysis.

\subsubsection{Local analysis in points with a horizontal tangent}

These points satisfy the system of two equations $F(h', h)=0$ and $\partial_{h'}F(h',h)=0$. For $h'\neq0$, the second equation is satisfied when $d^2\/\Fr^2\/(1+{h'}^2\/)^3 = (3\/\Bo\/h)^2$. From the last expression, we  find ${h'}^2$ as a function of $h$ and replace it in the equation $F=0$. Thus, we obtain a nonlinear equation depending only on $h$, but involving a cubic root operator. To get rid of this irrational function and deal only with polynomials, we introduce a new variable $Y$ such that $h=(d\Fr/3\Bo)Y^3$. Thence, ${h'}^2 = Y^2 - 1$, with $Y\geqslant1$. Eventually, we obtain the following polynomial equation of degree nine in $Y$ whose coefficients are polynomial functions of $\Fr$ and $\Bo$
\begin{equation}
  f(Y)\, \equiv\, \Fr^2\/Y^9\, -\, (3\Fr-2)\/\Fr\/\Bo\/Y^6\, +\, 9\/\Bo^2\/(1+2\Fr-2\Bo)\/Y^3\, + \,27\/\Bo^3\/Y^2\,-\,36\/\Bo^3\, = \,0.
\end{equation}
The appropriate tool to discuss the number of real roots of $f(Y)$ is the discriminant $D_1(\Fr,\Bo)$, which is a polynomial function in the variables $(\Fr, \Bo)$ and it generalises the well known discriminant of a quadratic equation in $Y$. The zero level set of $D_1$ is an algebraic curve $\D_1$ in the $(\Fr,\Bo)$-plane. It characterises the values of $(\Fr, \Bo)$ where $f(Y)$ experiences the collision of (real or complex) roots. For the algebraic definition and first properties of the discriminant see, \eg \cite{Nickalls1996} or Chapter~5 of the classical textbook \cite{Lang2002}. For multidimensional extensions, we refer to \cite{Gelfand1994}. The key property of our analysis is that the curve defined by the implicit equation $D_1(\Fr, \Bo) = 0$ divides the parametric $(\Fr,\Bo)$-plane into domains where $f(Y)$ has the same number of real roots. However, we would like to make the important observation that some branches of the discriminant locus $\D_1$ correspond to the collision of complex roots of $f(Y)=0$, without any visible effects on the real roots. In this case, these two domains have to be merged into a single one. These situations require a special post-processing procedure.

By definition of the $Y$ variable, its values of physical interest are necessarily greater or equal than one. Thus, the sub-domains of the $(\Fr,\Bo)$-plane where the additional condition $Y \geqslant 1$ is fulfilled by real roots are delimited by some ovals of the zero level-set $f_{\Fr, \Bo}(1)=0$. As above, it is possible that the zero level-set contains parasitic ovals which do not lead to any visible effects on the real roots, when one crosses their boundary. Eventually, we construct explicitly a partition of the rectangle $[0,2]\times[-1,2]$ into connected domains. In each cell, the number of local extrema of $h$ on $C_{\Fr,\Bo}$ is constant.

\subsubsection{Local analysis in points with a vertical tangent}

We follow the same methodology already employed above to study the points of the curves $C_{(\Fr, \Bo)}$ with a vertical tangent. They satisfy the system of two equations $F = 0$ and $\partial_h F = 0$. Computing $F(h,h') - h\partial_h F$, we eliminate the irrational terms. By introducing a new variable $Z$ such that ${h'}^2 = Z^2 - 1$, we arrive to the equation
\begin{equation}\label{eq:Z}
  \Fr(Z^2 - 1)\ =\ -3\,(\,h/d\, -\, 1\,)\left(\,2\,(h/d)^2\, +\, \Fr\, h/d\, +\, \Fr\,\right).
\end{equation}
We can also replace ${h'}^2$ by $Z^2 - 1$ in the irrational equation $F(h, h') = 0$ to obtain a new polynomial equation $G(h, Z) = 0$. Since now we are left with two polynomial equations (\ref{eq:Z}) and $G(h,Z) = 0$ for the variables $(h, Z)$, we can eliminate $h$ computing their resultant with respect to $h$. Thus, we obtain a polynomial $g_{\Fr,\Bo}(Z)$ of degree six in $Z$, which plays the same role that the polynomial $f_{\Fr,\Bo}(Y)$ in the previous section. Similarly, we compute the discriminant of the polynomial of $g$ with respect to the variable $Z$ to find a polynomial expression $D_2(\Fr,\Bo)$. As explained above, it allows to decompose the parameters space $(\Fr,\Bo)$ into cells where the number of points on the curve with a vertical tangent is the same. We also require that these points satisfy an additional condition $h > 0$, which physically means the absence of dry areas.

Finally, we take the intersection of two families of cells we constructed in this Section. As a result, we construct explicitly a cell decomposition of the rectangle $[0, 2] \times [-1,2]$. In each cell, the numbers of local extrema of $h$ and $h'$ on $C_{\Fr,\Bo}$ are preserved. Hence, we are able to classify all possible shapes of $C_{\Fr,\Bo}$ with respect to admissible paths connecting $A_1$ either to a point with $h' = 0$ or to a point with a vertical tangent.


\subsection{Partition with a fixed parameter}

Here, we describe the deformation of the curves when they undergo the continuous parameters change. For the sake of simplicity, we fix one parameter (say $\Bo = 0.32$) and vary the other one. The Froude number $\Fr$ will take the values in the segment $[0, 2.5]$ to produce a continuous family of curves $C_{\Fr, 0.32}$. As expected, the shape of the curves changes only for special values of the parameter $\Fr$. Thus, we obtain a decomposition of the segment $[0, 2.5]$ into a finite numbers of intervals where the shape is maintained. Below, we consider several typical behaviours. Keeping the notation of the previous section, we conduct the computations.

The discriminant $D_1(\Fr, 0.32)$ has four real roots, which are equal (approximately) to $-8.0793$, $0$, $0.96$, $0.966423$. In this list, we can recognise the two special values already detected above, \ie $\Fr=0$ and $\Fr=3\Bo=0.96$. The value $\Fr=-8.0793$ is outside the interval of physical interest. The discriminant $D_2(\Fr)$ has six real roots, equal approximately to $-8.8226$, $-8.0793$, $0$, $0.95678$, $0.966423$, $2.0382$. Again, we recognise the special value $\Fr=0$ along with two previously detected roots of $D_1(\Fr, 0.32)$. Moreover, we obtain a new value $\Fr=0.95678$, while the value of $\Fr=-8.8226$ is outside of the considered interval. During the local asymptotic analysis around the points where $h'=0$, we discarded the segment delimited by $(\Fr-1)(\Fr-3\Bo) < 0$, which corresponds to $[0.96, 1]$ in the present case. Finally, we have to analyse only the segments: $[0, 0.95678$], $[0.95678, 0.96]$, $[1, 2.0382]$ and $[2.0382, 2.5]$ along with their end points.

For each segment, it is sufficient to choose only one value for $\Fr$ (inside this interval) and plot the corresponding curve $C_{\Fr, 0.32}$. Here, we summarise our main findings:
\begin{enumerate}
  \item For $\Fr = 0$, there are neither vertical nor horizontal tangents. Therefore there are no non-trivial solutions.
  \item For $\Fr$ strictly between $0$ and $0.95678$, there are points with vertical as well as horizontal tangents to the curve, but their relative locations, with respect to the orientation, do not allow the existence of an admissible non-trivial solitary wave. See Figure~\ref{fig:CC02,032}(\textit{a}) for an illustration.
  \item The value of $\Fr = 0.95678$ corresponds to the collision of complex roots and it does not bring anything for the real curve. Consequently, this boundary point has to be removed from the present discussion.
  \item For $\Fr = 0.96$, which is a special value since $\Fr = 3\Bo = 0.96$, there is only one admissible path to the point with a vertical tangent; see Figure~\ref{fig:CC02,032}(\textit{b}) for the illustration. It corresponds to a single peakon-like wave. It has an additional interesting property, since it decreases only algebraically to infinity (all other solutions considered herein above had an exponential decay).
  \item For $\Fr$ lying strictly between $0.96$ and $1$, as noticed earlier, there are only one isolated point at $(h=1, h'=0)$. Therefore there are no non-trivial solutions.
  \item For $\Fr = 1$, which is also a special critical value, the curve $C_{1, 0.32}$ admits a cusp at $(h=1, h'=0)$ and there are neither vertical nor horizontal tangents. Therefore, there are no non-trivial solutions in this case; see Figure~\ref{fig:CC1,032}(\textit{a}).
  \item For $\Fr$ strictly greater than $1$ and smaller than $2.0382$, there are points with a vertical as well as horizontal tangents to the curve $C_{\Fr, 0.32}$; see Figure~\ref{fig:CC1,032}(\textit{b}). Notice that the loop of this curve resembles the one analysed in our first example of Section~\ref{sec:Ex1}. The only difference consists in the loop orientation (upward). So, for this range of values there is a regular solitary wave with a smooth crest. By including discontinuous patches, we can also construct admissible peakons with a finite number of crests, as illustrated in Figure~\ref{fig:ex1}(\textit{b}).
  \item The value of $\Fr = 2.0382$ is not relevant for our discussion. 
\end{enumerate}


\subsection{Types of behaviours}

For our final classification, it is worthwhile to give names to the different kind of qualitative behaviours we already encountered. Let us denote by a $0$, $I$, $II$, the types which appear generically (on open subdomains of the parameters space), and by $E_1$, $E_2$, $E_3$ the more extreme types which appear on the discriminants components (lines or curves). So, let us call
\begin{itemize} 
  \item Type 0, the kind of situation, like in the first or second previous items, where no regular or generalised solitary wave may occur, besides the trivial one $h=d$.
  \item Type I, the kind of situation, like in Example 1, where there are a regular wave with a trough and also generalised peakon solutions with one or several angles.
  \item Type II, the kind of situation, like in the previous item (vii), where there are a regular wave with a crest and also generalised peakon-like solutions with one or several angles.
\end{itemize}
Also, let us name the first kind of extremal situation we already encountered:
\begin{itemize} 
  \item Type $E_1$, the kind of situation, like in the third previous item, $\Fr = 3\Bo$, but $\Fr \neq 1$, with a single peakon, \ie it admits a symmetric single angular solitary wave. It also has an algebraically decrease to infinity. 
\end{itemize}
We will see that the very extremal case $\Fr = 3\Bo$ and $\Fr = 1$, admits the same type of solutions.

\begin{figure}
  \centering
  \subfigure[]{\includegraphics[width=0.48\textwidth]{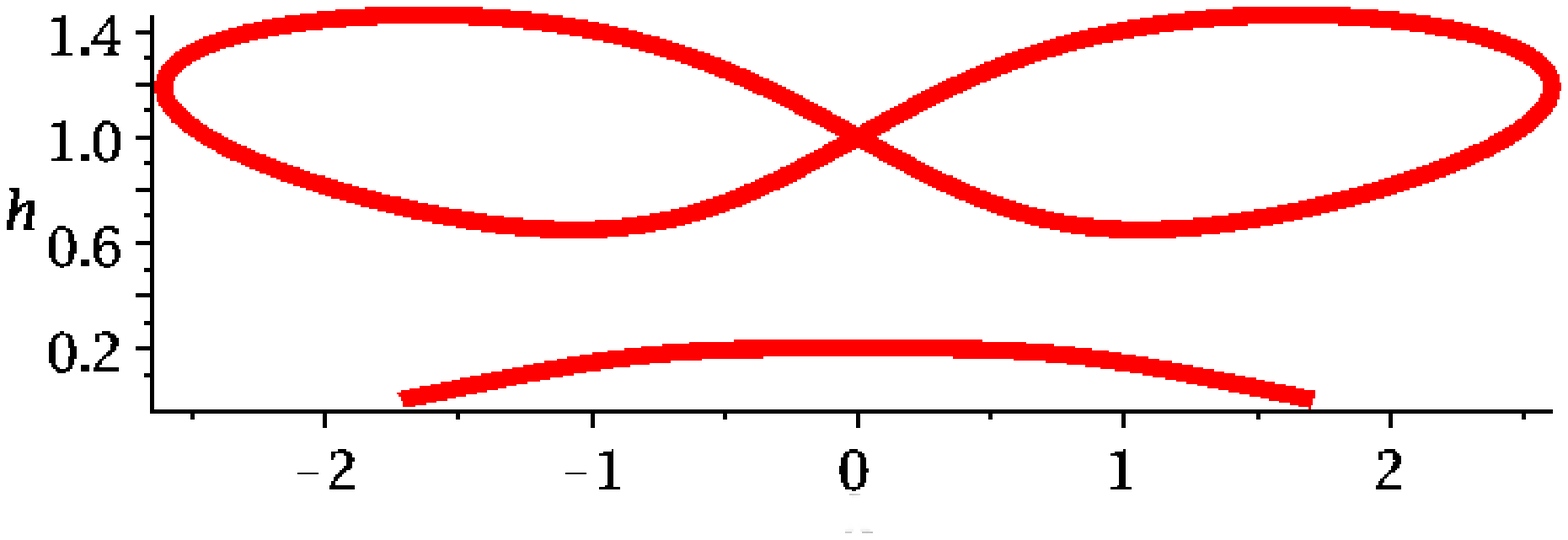}}
  \subfigure[]{\includegraphics[width=0.48\textwidth]{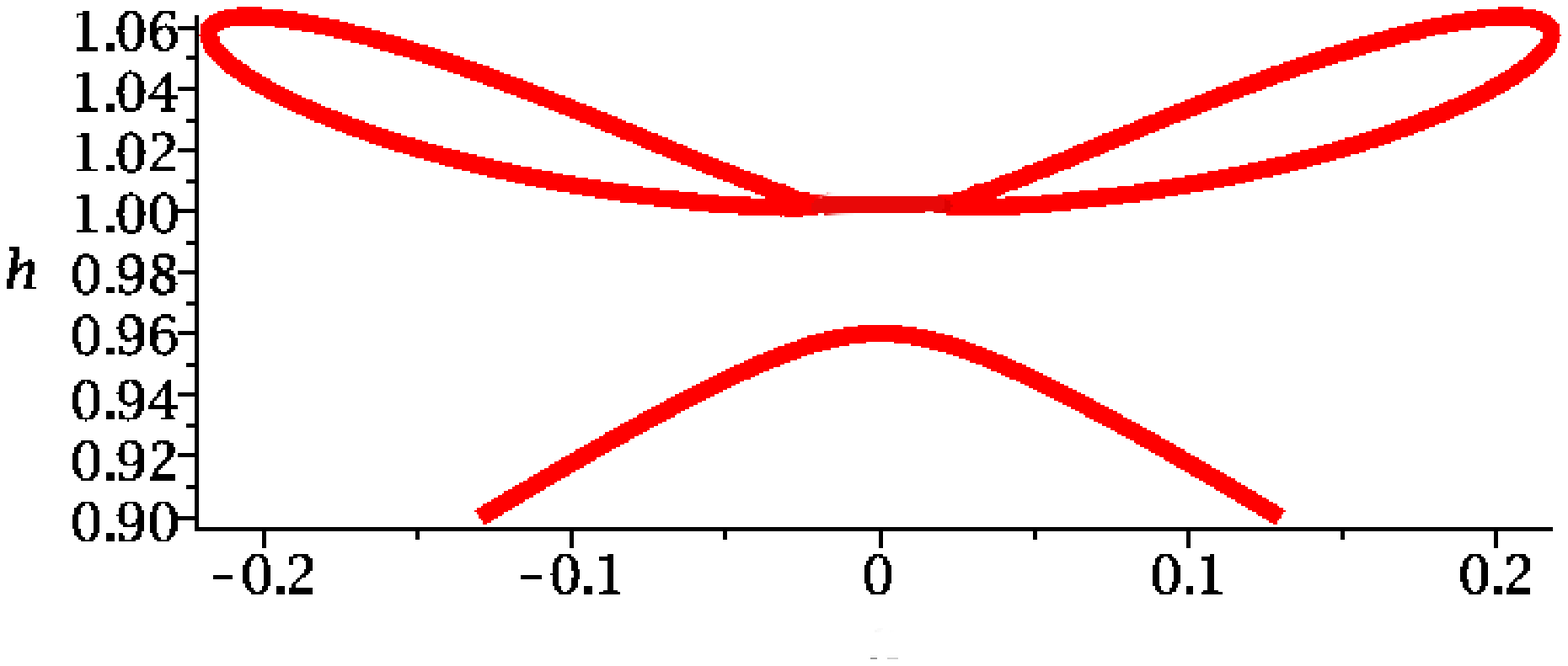}}
  \caption{\small\em Algebraic curve in the phase space (a) $C_{0.2,\ 0.32}$ and (b) $C_{0.96,\ 0.32}$.}
  \label{fig:CC02,032}
\end{figure}

\begin{figure}
  \centering
  \subfigure[]{\includegraphics[width=0.48\textwidth]{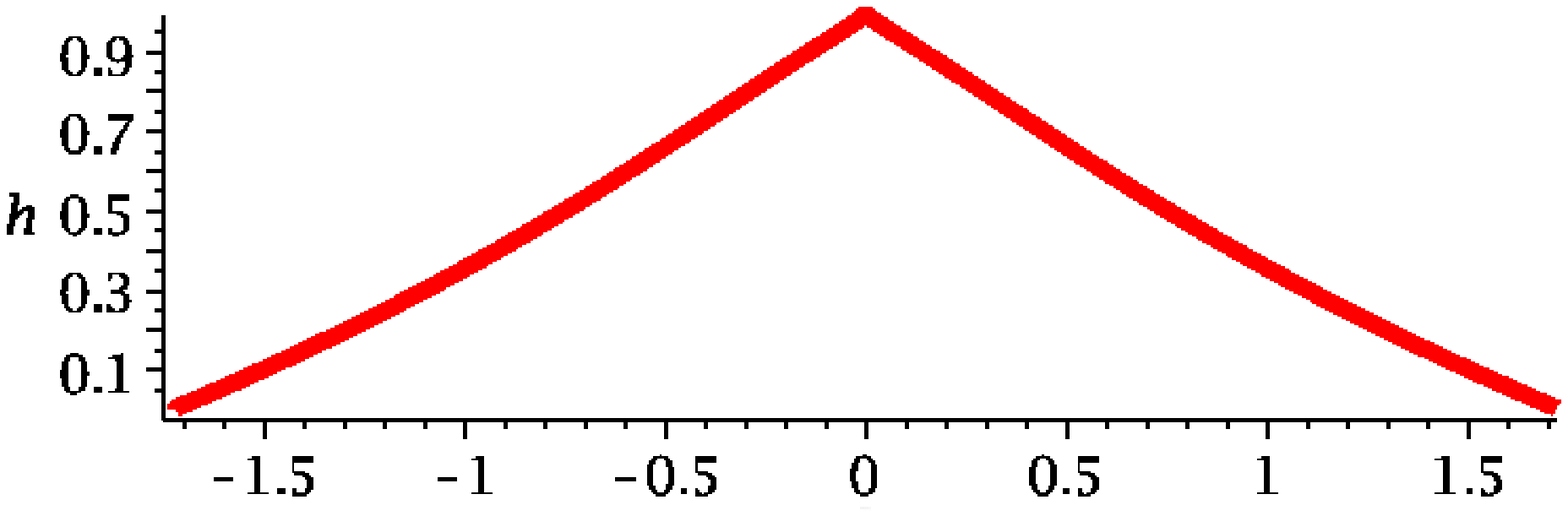}}
  \subfigure[]{\includegraphics[width=0.48\textwidth]{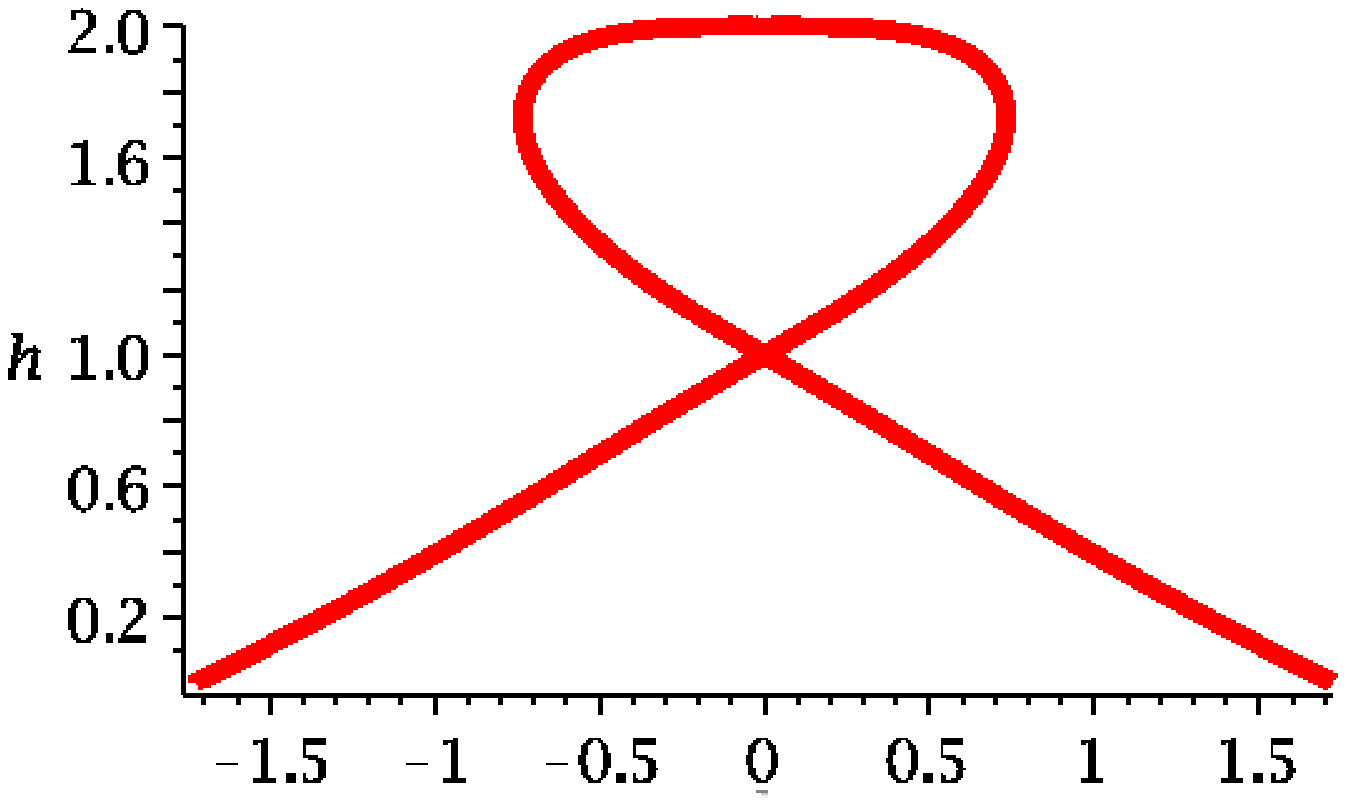}}
  \caption{\small\em Algebraic curve in the phase space (a) $C_{1,\ 0.32}$ and (b) $C_{2,\ 0.32}$.}
  \label{fig:CC1,032}
\end{figure}


\subsection{Configuration space partition}

\begin{figure}
  \centering
  \subfigure[]{\includegraphics[width=0.48\textwidth]{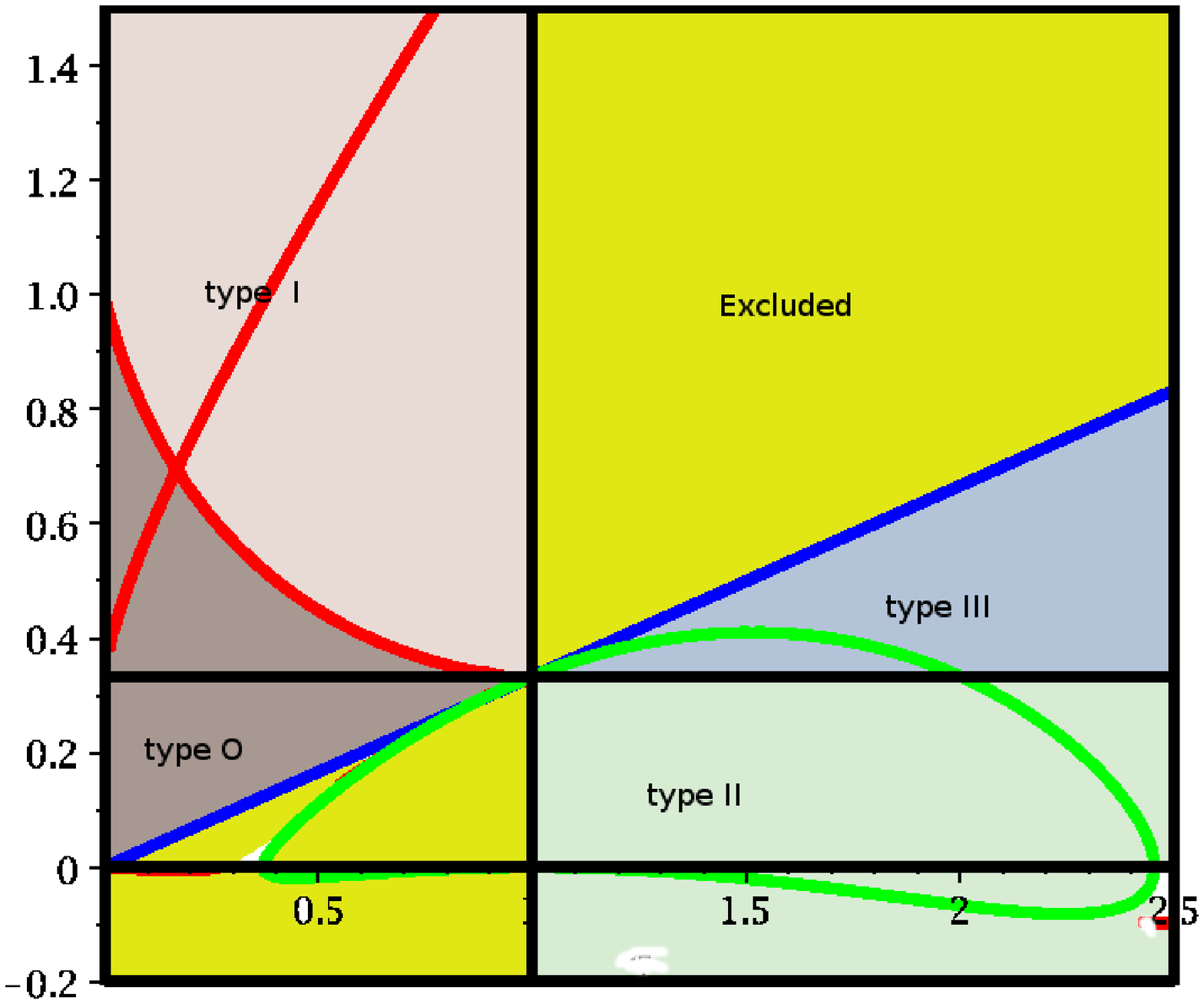}}
  \subfigure[]{\includegraphics[width=0.48\textwidth]{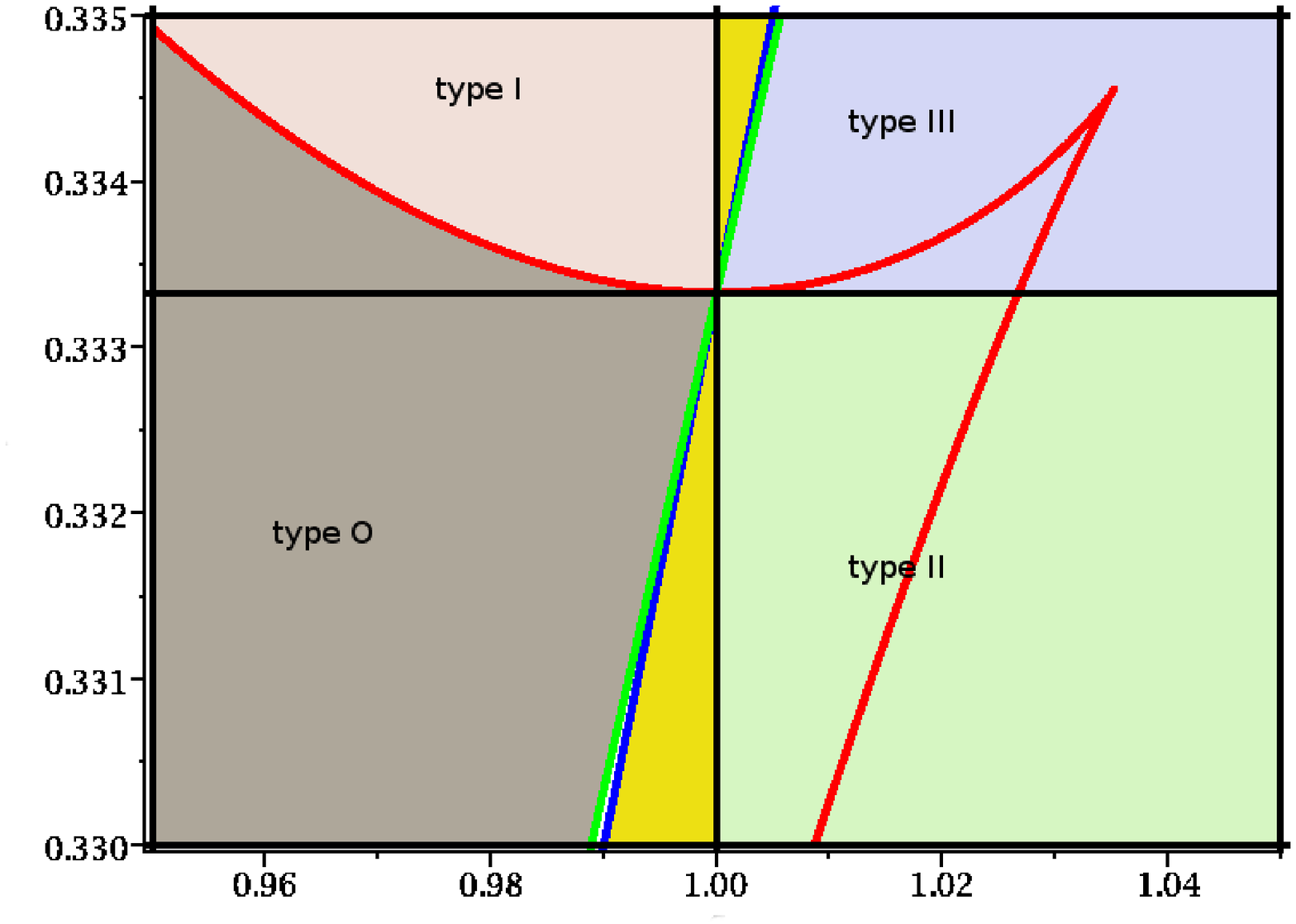}}
  \caption{\small\em Discriminant loci (a) and (b) a zoom on region around the point $(1, \frac13)$.}
  \label{discrim}
\end{figure}

In this section, we consider both the parameters $(\Fr, \Bo)$ as variables which are free to take any values from the rectangle $[0, 2]\times[-1, 2]$. Our aim is to construct a cellular decomposition of this domain, each cell containing a different type of behaviour of the solution. For this purpose, we employ the computer algebra system \textsc{Maple}. More specifically, using \textsc{Maple} commands $discrim()$ and $factor()$, we can show that the discriminant polynomial $D_1$ can be decomposed into the product of a square $(\Fr-3\Bo )^2$ and another polynomial of degree $10$ that we denote by $D(\Fr,\Bo)$. The zero locus of $D$ (shown in red) together with the special lines $\Bo=1/3$, $\Bo=\Fr/3$ and $\Fr=1$ are shown in the figure~\ref{discrim}. A zoom on a region of interest around $\Bo=1/3$ and $\Fr=1$ is provided in the figure~\ref{discrim}({\it b}).

Similarly, we decompose the other discriminant polynomial $D_2(\Fr,\Bo)$. As a result, we find the product of the same polynomial $D(\Fr,\Bo)$ by powers of $\Fr$, powers of $\Bo$ and by the cube of another polynomial that we denote by $D_3(\Fr,\Bo)$. The zero locus of $D_3(\Fr,\Bo)$ is shown in green in Figure~\ref{discrim}.

Figure~\ref{discrim} shows only a preliminary partition of the area under consideration. As it was illustrated in the previous section on a simple 1D example (with the variable $\Fr$ number), some boundaries are artificial in the real domain. Consequently, below we analyse separately each case.

The line in the parameters plane and defined by the linear relation $\Fr=3\Bo$ gives rise to a double root at a fixed value $Y = 1$. In other words, it is the double point at $h = 1$ and $h' = 0$.

\subsubsection{Additional types and classification}

Considering each of the delimited domains, allow to classify all the possible curves $C_{\Fr,\Bo}$, by the relative location of their points with horizontal tangent, vertical tangent, singularity. For each such type of curve, we analysed all the possible admissible paths (with respect to the orientation) departing from $A_1=$ and arriving to a point with a vertical tangent, or to a point of the line $h'=0$. To each of them a regular or an angular solitary wave is associated.

\begin{figure}
  \centering
  \subfigure[]{\includegraphics[width=0.49\textwidth]{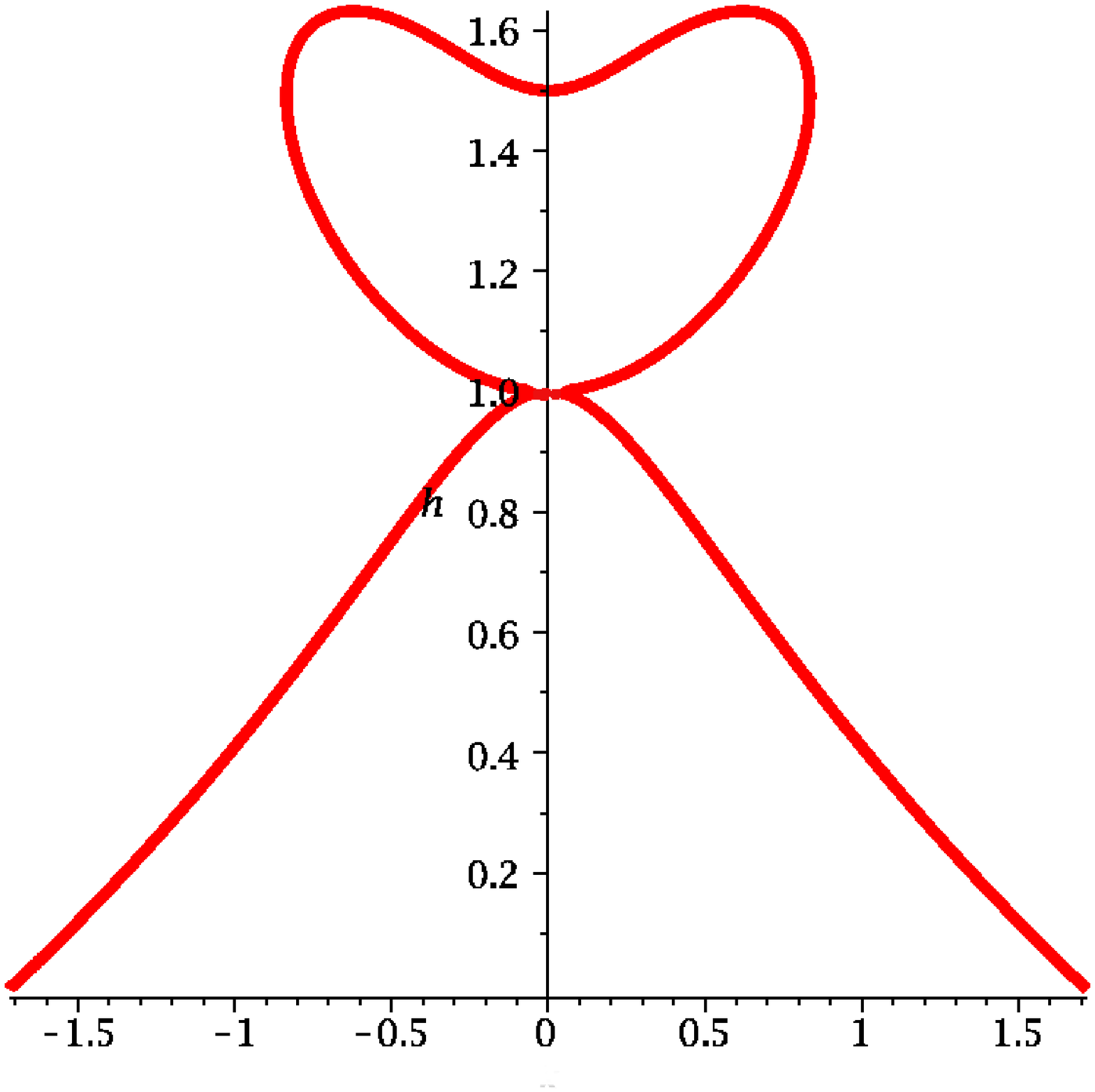}}
  \subfigure[]{\includegraphics[width=0.49\textwidth]{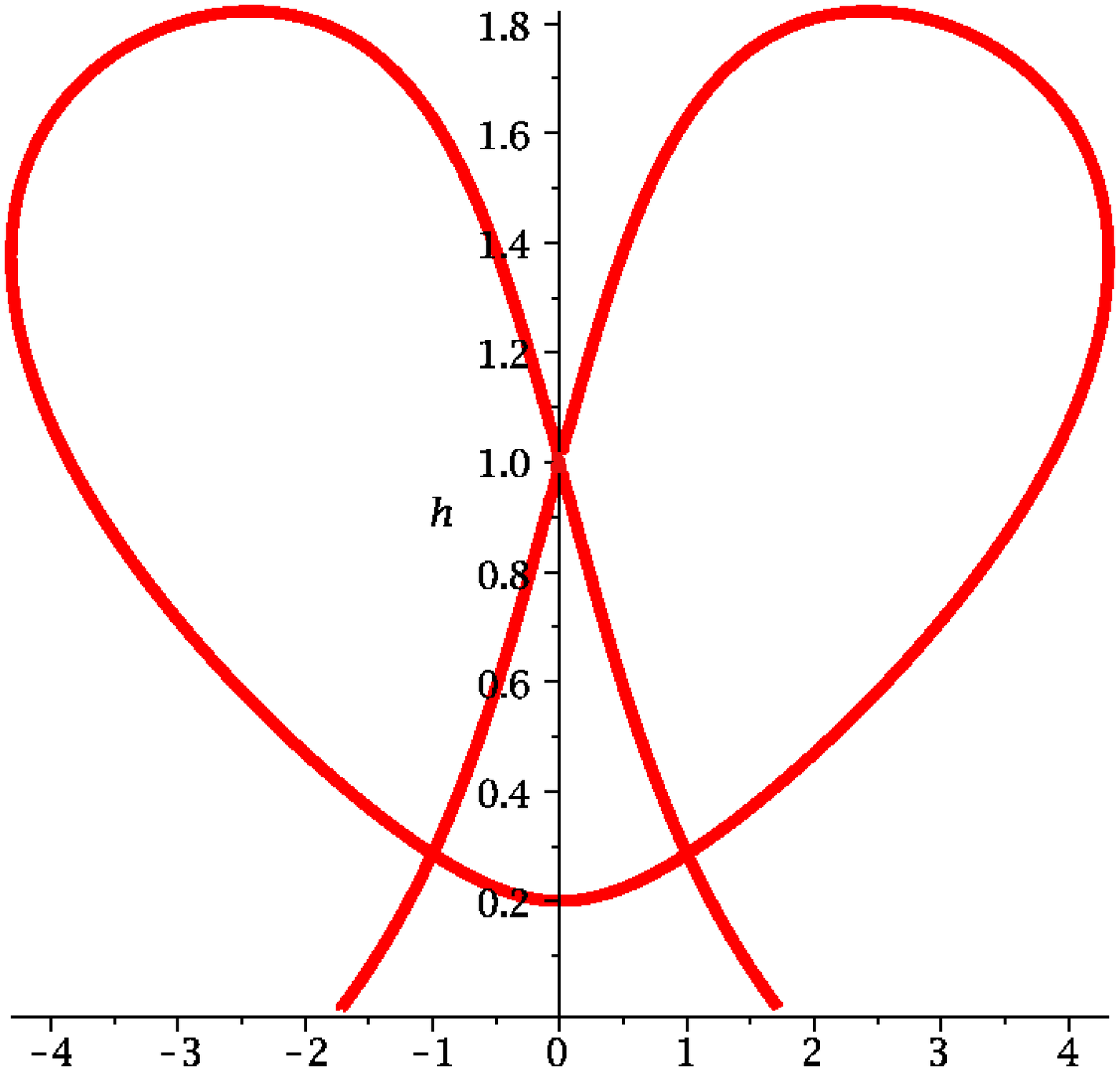}}
  \caption{\small\em (a) A semi-extremal and extremal (b) curves.}
  \label{fig:extremal}
\end{figure}

We found only two types of regular solitary waves: a crest when $\Fr>1$, that we called type II, and a through when $\Fr<1$, that we called type I. We have to introduce a new ``generic'' type.
\begin{itemize}
  \item Type III corresponds to the open subset of the parameter space with $\Fr > 1$, $\Fr > 3\Bo$ and $\Bo > 1/3$, where there are no regular wave but generalised peakon-like solutions with one angle like illustrated in the Figure~\ref{fig:extremal}(\textit{a}).
\end{itemize}

\begin{figure}
  \centering
  \subfigure[]{\includegraphics[width=0.49\textwidth]{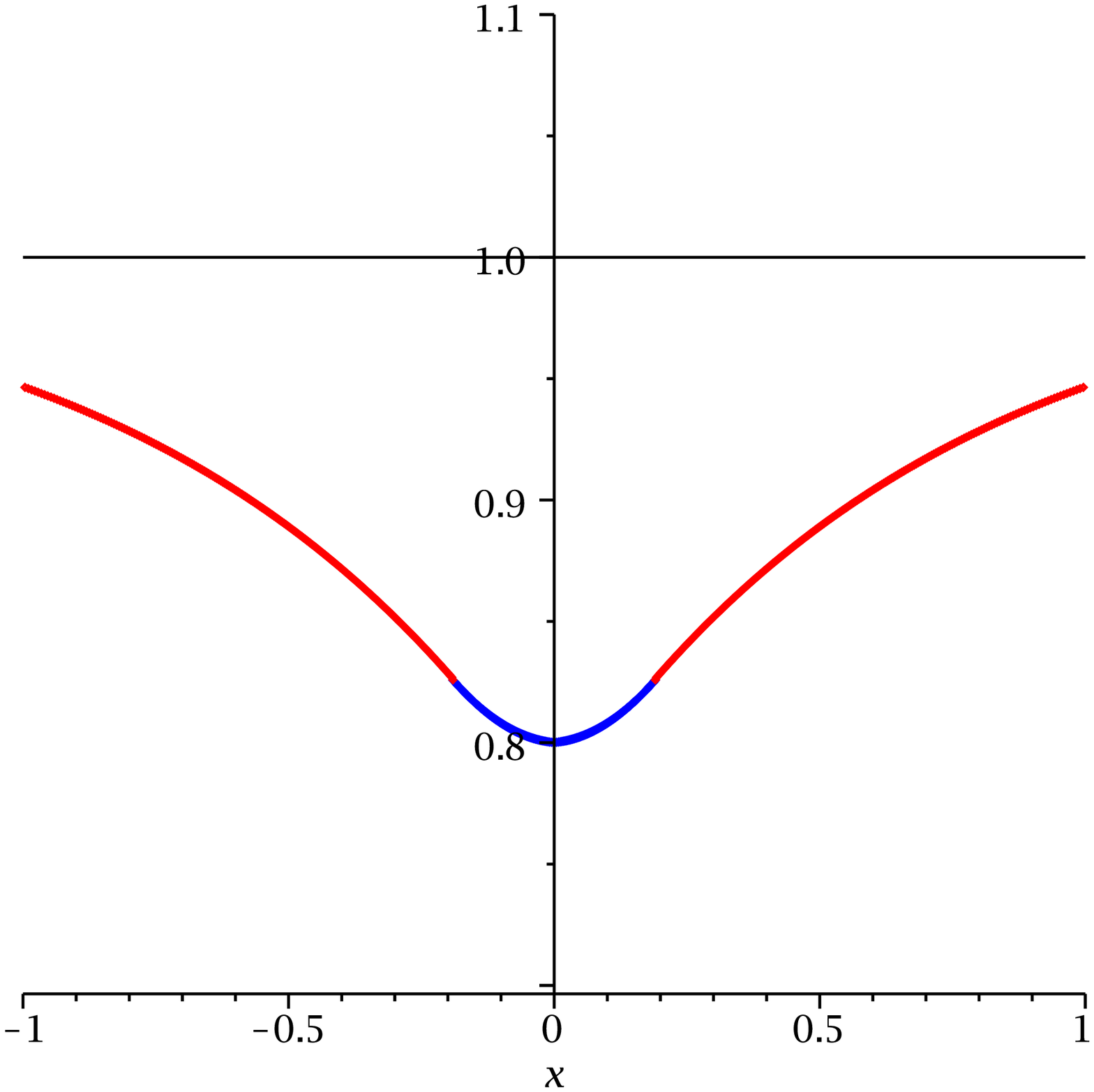}}
  \subfigure[]{\includegraphics[width=0.49\textwidth]{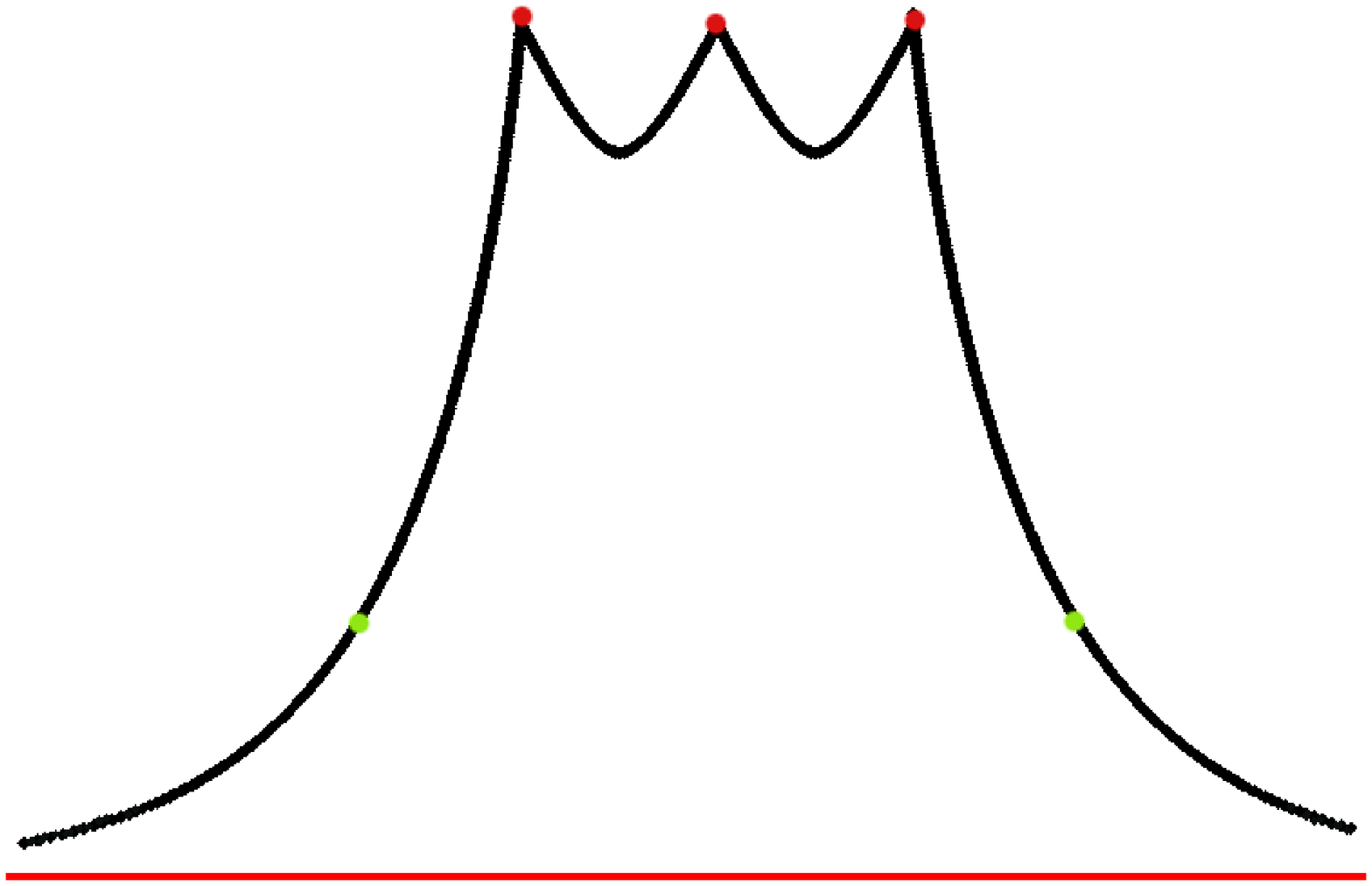}}
  \caption{\small\em (a) A weakly regular solitary wave. (b) A weak multiangular crested wave.}
  \label{fig:weakRegular}
\end{figure}

Besides type $E_1$, we found the following extremal types:
\begin{itemize}
  \item Type $E_2$. The limiting case where $(\Fr,\Bo)$ belongs to a branch of the discriminant locus, \eg when $(\Fr=0.2,\Bo=0.66)$ approximately, we obtain the curve shown in the figure \ref{fig:extremal}(b). There is a connected path but with an angle going from $A_1$ to a lower point on $h'=0$. It gives rise to a weak regular wave shown in Figure~\ref{fig:weakRegular}({\it a}), which has two points where the regularity is only $C^1$ (we represented them by changing the colours of the branches).  
  \item Type $E_3$. A tiny branch of the discriminant with $\Fr>1$ gives rise to another interesting type of ``semi extremal curve'', \eg $C_{1.01,0.3334121494}$, shown in Figure~\ref{fig:extremal}(\textit{a}) and corresponding to (multi) angular weak crests (Figure~\ref{fig:weakRegular}{\it b}). The angles are indicated with a red disk and the weak defect of regularity are indicated with a green disk.
\end{itemize}

\section{Conclusions and future work}
\label{sec:concl}

In this paper, we addressed the problem of constructing weak solitary wave solutions using phase plane analysis and applying algebraic geometry techniques. The method is illustrated on a weakly dispersive fully nonlinear capillary-gravity waves model equations, namely the \textsc{Serre--Green--Naghdi} (SGN) equations. These equations are not trivial but yet tractable, so the power of our approach could be illustrated. 

After deriving the governing SGN system of equations, we restricted our attention to a particular case of travelling solitary wave solutions. This study contains a complete phase space analysis of all admissible solutions. Among them we found the classical infinitely smooth solitary waves of elevation ($\Fr > 1$) and of depression ($\Fr < 1$). However, by including into consideration the discontinuous paths on the phase curves, we were able to construct peakon-like solitary waves with angular points at the crests (respectively, the troughs). These `peakons' are valid mathematical solutions to the ODE describing steady waves. Consequently, this work can be considered as a continuation of previous important studies on peaked travelling waves in some shallow water model equations (\eg \textsc{Camassa--Holm}). We are aware that these solutions are of little physical applicability, since the model is pushed towards its limits. (The smooth weak solution of Figure \ref{fig:weakRegular}({\it a}) may be physically sound, however.) Nevertheless, we believe that this work can suggest directions for improving the model. Also, singular (\ie angular) solutions of fluid flows in presence of surface tension exist physically \cite{Eggers2009}, so our analysis could be applied to model and study these phenomenae.

We reduced the characterisation and classification of all admissible solutions (regular or peakon-like solitary waves) to a geometric problem. Namely, for each values of a pair $(\Fr, \Bo) \in [0,2]\times[-1,2]$, we found if it exists on the curve $C_{(\Fr, \Bo)}$ an ``admissible'' path starting and ending at the point $A_1$ (corresponding to $h'=0, h=d$, a boundary condition satisfied by solitons). Our strategy to solve this geometric problem was to first provide a complete classification of possible phase curve topologies using some advanced computer algebra techniques, and then discard phenomena due  to  branches not connected to $A_1$. The algebraic methods were explained here in details with illustrative examples. This study is a successful example of the application of some methods of effective algebraic geometry to the qualitative analysis of ODEs stemming from Fluid Mechanics problems. Despite the purely mathematical interest, we were able also to find some new types of solitary waves. For example, one can mention the weakly singular solitary wave (with a jump in the second derivative) along with the wave decaying algebraically at the infinity.

With our analysis, we pushed the SGN model towards its limits, since the Laplace capillarity law does not apply when the derivatives of the elevation jumps at a peak. Nevertheless, one observes that when we zoom out some regular sharp profile of solitary wave, it looks like a peakon. Then, for a fixed  pair of parameters $(\Fr,\,\Bo)$, since we deal with an approximate model, we can interpret such a peakon solution as a warning. A sign for the possible existence of some sharp regular solitary wave of the exact model, with a near-by pair of parameters. This argument is emphasised when the phase plane curve corresponding to a peak is similar to a phase plane giving rise to a regular solitary wave; for example, for the curves of type III (in our classification), which are deformations of curves of type II.

In future works, we first plan to study in a similar way the families of periodic travelling wave solutions. This problem is sensibly more complicated, since its formulation involves two additional parameters (integration constants) depending implicitly on the solution. Thus, this problem leads us to work in the four-dimensional parameter space. Secondly, another research direction consists in looking for an augmented shallow water model with possibly some additional physical parameters. These extra degrees of freedom would be used to represent singular solutions as a limit of smooth ones. The main motivation for this consists in the fact that it is already the case for the full \textsc{Euler} (limiting \textsc{Stokes} wave) and the \textsc{Camassa--Holm}-type equations.


\subsection*{Acknowledgments}
\addcontentsline{toc}{subsection}{Acknowledgments}

The authors would like to acknowledge the support from CNRS under the PEPS 2015 Inphyniti programme and exploratory project FARA. D.~\textsc{Dutykh} would like to thank the hospitality of the Laboratory J. A.~Dieudonn\'e and of the University of Nice -- Sophia Antipolis during his visits to Nice.


\addcontentsline{toc}{section}{References}
\bibliographystyle{abbrv}
\bibliography{biblio}

\end{document}